\documentclass{article}
\usepackage{slashed,verbatim}
\usepackage[numbers,sort&compress]{natbib}
\usepackage{amsmath}

\newcommand{\Rmnum}[1]{\expandafter\@slowromancap\romannumeral #1@}
\makeatother
\usepackage{amssymb}
\usepackage{float}
\usepackage{comment}
\usepackage{appendix}
\usepackage{tikz-feynman}
\usepackage{graphicx}
\usepackage{subcaption}
\usepackage{dcolumn}
\usepackage{bm}
\usepackage[colorlinks=true,linktocpage=true,
linkcolor=blue,citecolor=blue]{hyperref}
\usepackage[a4paper]{geometry}
\newcommand{\nd}{\noindent}
\newcommand{\be}{\begin{eqnarray}}
\newcommand{\ee}{\end{eqnarray}}

\begin{document}
\large
\title{\bf{Chirality dependence in charge and heat transport in thermal QCD}}
\author{Pushpa Panday\footnote{pushpa@ph.iitr.ac.in}~~and~~Binoy Krishna
Patra\footnote{binoy@ph.iitr.ac.in}\vspace{0.1in}\\
Department of Physics,\\
Indian Institute of Technology Roorkee, Roorkee 247667, India}
\maketitle

\begin{abstract}
As the strength of the magnetic field ($B$) becomes weak, novel phenomena, 
similar to the Hall effect in condensed 
matter physics, emerges both in charge and heat transport in
a thermal QCD medium with a finite quark chemical potential ($\mu$).
So we have calculated the transport coefficients 
in a kinetic theory within a quasiparticle 
framework, wherein we compute the effective mass of 
quarks for the aforesaid medium in a weak magnetic field 
(B) limit ($|eB|<<T^2$; T is temperature) by the perturbative thermal 
QCD up to one loop, which depends on $T$ and $B$ differently to 
left- (L) and right-handed (R) chiral modes of quarks, lifting 
the prevalent degeneracy in L and R modes in strong magnetic field 
limit ($|eB|>>T^2$). Another implication of weak 
$B$ is that the transport coefficients assume a tensorial structure: 
The diagonal elements represent the usual (electrical and thermal) 
conductivities: $\sigma_{\rm Ohmic}$ and $\kappa_0$ as the 
coefficients of charge and heat transport, respectively 
and the off-diagonal elements denote their Hall counterparts: 
$\sigma_{\rm Hall}$ and $\kappa_1$, respectively. 
It is found in charge transport that the magnetic field acts on 
L- and R-modes of the Ohmic-part of electrical conductivity in 
opposite manner, {\em viz.} $\sigma_{\rm Ohmic}$ for L- mode decreases 
and for R- mode, increases with $B$ whereas the Hall-part $\sigma_{\rm Hall}$
for both L- and R-modes always increases with $B$. 
In heat transport too, the effect of the magnetic field on the usual thermal 
conductivity ($\kappa_0$) and Hall-type coefficient ($\kappa_1$) in both 
modes is identical to the abovementioned effect of $B$ on charge 
transport coefficients.  

We have then derived some coefficients from the above transport 
coefficients, {\em namely} Knudsen number ($\Omega$ is the ratio of 
the mean free path to the length scale of the system)
and Lorenz number in Wiedemann-Franz law. The effect of $B$ on $\Omega$
either with $\kappa_0$ or with $\kappa_1$ for both modes are identical to
the behaviour of $\kappa_0$  and $\kappa_1$ with $B$. The value of 
$\Omega$ is always less than unity for the entire temperature range,
validating our calculations. Lorenz number ($\kappa_0/\sigma_{{\text{Ohmic}}}T$) and Hall-Lorenz number ($\kappa_1/\sigma_{{\text{Hall}}}T$) for L-mode increases and for R-mode decreases with magnetic
		field. It also does not remain constant with temperature hence violating the Wiedemann-Franz law.
\end{abstract}
\section{INTRODUCTION}
 Quark-gluon plasma (QGP) is the deconfined phase of 
 quarks and gluons which is believed to have existed 
 in the early universe, about $10^{-5}$sec after the 
 cosmic Big Bang and at the core of superdense stars 
 such as neutron stars and quark stars. Experiments
  at European Council for Nuclear Research (CERN), 
  Relativistic Heavy Ion Collider (RHIC), Brookhaven
   National Laboratory (BNL) and Large Hadron Collider 
   (LHC) have been successful in creating QGP in 
 colliders \cite{Newforms}. It is also established that
  a magnetic field, whose magnitude varies from 
  $|eB|$ = 0.1 $m_{\pi}^2$ for SPS energy to 
  $|eB|$ = 15 $m_{\pi}^2$ for LHC, is also produced 
  during non-central heavy ion collisions 
  \cite{skokov, kharzeev, fukushima}. The strength of 
  this magnetic field is strong during the initial 
  stages of QGP but it decays very fast with time. 
  The life-time of magnetic field in a charged 
  medium, however, gets enhanced due to the charge properties of the medium 
   \cite{tuchin, k_tuchin, Tuchin, Mclerran}. As the non vanishing magnetic field can affect the 
   evolution of strongly interacting matter significantly
   \cite{a_das, tuchin_photon_deacy, greif, m_greif, puglisi, a_puglisi, cassing, steinert}, therefore the detailed study of its effects on transport 
   phenomena \cite{s_rath, debarshi}, thermodynamical 
   behaviour \cite{suba, karamkar} of quark-gluon plasma, 
   dilepton production from QGP \cite{K_Tuchin, peshier, mamo} 
   has been done.
  Further, the bulk evolution of QGP matter via relativistic hydrodynamics has been described successfully, which gave satisfactorily the collective flow of the created matter detected in experiments \cite{heinzl,schenke,niemi}. 
  The small ratio of shear viscosity 
  to the entropy density ($\eta/s$) of strongly interacting plasma agrees
   well with the lower bound of $\eta/s = \frac{1}{4\pi}$, where $\hbar = 1, k_B = 1$, obtained using AdS/CFT correspondence 
   \cite{kovtun} hence, validates the use of hydrodynamical model of QGP \cite{romatschke, teaney, huovinen, baier, heinz}. Hydrodynamical description of QGP evolution after heavy-ion collision requires, stating various transport coefficients, which can be interpreted as medium's response to various perturbations.
   
  We study the charge and heat transport coefficients which also
    plays an important role in the hydrodynamical
  description of strongly interacting matter 
   \cite{denicol, s_ghosh, g_kadam}. The topological 
   effects induced by magnetic field can be quantified 
   using electrical conductivity and plays a crucial 
   role in the study of chiral magnetic effect 
   \cite{Fukushima}, which is a signature of $CP$ violation 
   in the strong interaction. Dilepton and photon production 
   rates are used to probe the thermalized strongly 
   interacting matter because they hardly interact 
   with the hadrons in region of hot and dense matter 
   and hence carry an information about the early 
  stage of heavy ion collisions. Electrical conductivity ($\sigma_{\text{el}}$) can be used for phenomenological 
  studies of heavy ion collisions \cite{haglin}. 
  Another key transport coefficient is thermal 
  conductivity of QGP medium, which measures the 
   transport of heat due to temperature gradient 
   in the medium. The hydrodynamical equilibrium 
   of the system can be determined using Knudsen 
   number, which is the ratio of mean free path to 
   the characteristic length of the medium. The mean
   free path ($\lambda$) is related to the thermal 
 conductivity ($\kappa$) as $\lambda = 3\kappa/(\mathrm{v} C_v)$, 
 where $\mathrm{v}$ is the relative velocity of 
 quark and $C_v$ is the specific heat at constant 
 volume. Further, the relative behaviour of $\kappa$ 
 and $\sigma_{\text{el}}$ can be understood in terms 
 of Wiedemann-Franz law, which states that ratio, 
 $\kappa/\sigma_{\text{el}}$, of the thermal to 
 electrical conductivity is directly proportional 
 to the temperature, with proportionality constant 
 being roughly same for all metals. The ratio 
 $\kappa/(\sigma_{\text{el}}T)$ is known as 
 Lorenz number ($L$), which is independent of 
 temperature and depends on fundamental constants 
 for all metals \cite{ashcroft}. However, the 
 violation of Wiedemann-Franz law has been observed 
 in many systems, such as hydrodynamic electron 
 liquid \cite{principi}, high temperature superconductors 
 \cite{hill}, Luttinger liquid \cite{garg}, 
 strongly interacting QGP medium \cite{mitra} and 
 hot hadronic matter \cite{rath}. Hence, it would 
 be interesting to study the Wiedemann-Franz 
 law in our system of interest.  
 
 In the present work, we have explored the effect of (a) weak magnetic field and (b) baryon asymmetry, in charge and heat transport phenomena. The weak and strong magnetic field limit can be understood from the relativistic dispersion relation of a fermion of mass $m$ in a uniform magnetic field ($\textbf{B}= B\hat{z}$): 
\begin{equation}
	E_n^2 = p_z^2 + m^2 + 2nqB.
\end{equation} 
Here, $n= 0, 1, 2, \cdots$ denotes the Landau levels. 
 The probability of fermions getting thermally excited to higher Landau levels 
is exponentially suppressed as $\exp\left(\frac{-\sqrt{qB}}{T}\right)$ \cite{k_fuku}. i) In strong magnetic field (SMF) limit, $\sqrt{qB}>>T$, so the fermions occupy only the lowest Landau level (n=0). This is known as LLL approximation. 
ii) If $\sqrt{qB}<<T$, then fermions 
can occupy higher Landau levels.
This implies that the thermal energy is much larger than 
the energy level spacing ($\sim \sqrt{qB}$) so that $T$ 
can excite fermions into the excited states, which justifies calling the condition $qB<<T^2$, the weak magnetic field (WMF) limit. The transport coefficients can be calculated in strong and weak magnetic field
 using different approaches/models, \textit{viz},
  NJL model \cite{marty, lang, ghosh}, Chapmann-Enskog 
  approximation \cite{wiranata, plumari, s_mitra}, 
  the correlator technique using Green-Kubo 
  formula \cite{S_ghosh, a_haruty, demir, sarthak}, 
  effective fugacity model 
  \cite{m_kurian, kurian,m.kurian,gow}, 
  lattice simulation \cite{s_gupta, aarts, ding}.
 However, we have used the kinetic theory approach 
 by solving the relativistic Boltzmann transport 
 equation. The calculation of transport coefficients 
 using kinetic theory has been done \cite{S_rath, debarshi, salman} 
 in presence of strong magnetic field ($q_fB>>T^2, m_f^2$) , 
 where $q_f$ and $m_f$ are the electric charge and 
 mass of quark for $f$th flavor. In a strongly
  magnetized medium, the motion of charged particle 
  is restricted to the $1+1-$dimensional Landau level 
  dynamics, where quark momentum is along the direction 
of magnetic field. In presence of weak magnetic field, 
however, temperature is the dominant energy scale 
($T^2>q_fB>m_f^2$) and effect of magnetic field comes 
through the cyclotron frequency ($\omega_c$). In 
contrast to the case of a strong background magnetic 
field,  motion of charges is no longer restricted 
to be along the direction of magnetic field, which 
gives rise to `transverse' responses. This can also 
be understood via the tensor structure of the transport 
coefficients at the two magnetic field strength 
regimes. In the case of strong magnetic field, 
the coefficient matrix is diagonal, whereas in 
the presence of a weak magnetic field, off-diagonal 
elements also manifest. The off diagonal elements
 are represented by $\sigma_{{\text{Hall}}}$ and 
 $\kappa_{1}$ in the case of electrical and 
 thermal conductivities respectively. This is 
 corroborated by the fact that there is no 
 $\sigma_{{\text{Hall}}}$ and $\kappa_{1}$ in 
 the case of strong magnetic field. Furthermore, 
 $\sigma_{{\text{Hall}}}$ and $\kappa_{1}$ vanish
 even in the presence of a weak magnetic field if 
 the chemical potential, $\mu$ is zero \cite{arpan}. 
 The role of interaction among partons is incorporated 
 using quasiparticle description of partons, where vacuum 
masses of partons are replaced by medium generated masses. 
The medium generated mass is calculated from the pole of 
propagator, obtained through perturbative thermal QCD in 
the presence of background weak magnetic field. 
In some previous studies, authors have incorporated 
the pure thermal medium mass of quarks in the 
computation of transport coefficients \cite{lata, arpan}, 
whereas we have used the thermally generated mass 
with magnetic field correction. The dispersion 
relation of quasiparticles in weak magnetic field 
give rise to four collective modes two from left-handed 
and two from right-handed modes. Various properties 
of dispersion relation have been discussed in
\cite{aritra,weldon}. The degeneracy in left- and 
right-handed chiral modes of quarks is lifted due to their different mass in 
the presence of weak magnetic field, which is in 
contrast to the case of strong magnetic field. The system can be either in left-handed mode or right-handed mode, hence 
the medium generated masses for left- and 
right-handed chiral modes of quarks have been taken into account separately 
for the estimation of transport coefficients under both modes. We further studied the physical behaviour of system using the aforementioned transport coefficients via Knudsen number and Wiedemann-Franz law for both modes separately.\\

 The paper is organised as follows: in Sec. 
 \ref{quasiparticle}, we discuss the quasiparticle 
 model of partons and hence evaluate the medium 
 generated mass. We use this mass as an input to 
incorporate the interactions among partons, in our
 calculation of transport coefficients. In Sec. 
 \ref{charge_transport} and Sec. \ref{thermal_transport}, 
we discuss the computation of charge and heat 
transport coefficients using kinetic theory within 
the relaxation time approximation. 
In Sec.\ref{results}, we present and discuss 
the results for Ohmic and Hall conductivity, 
thermal and Hall-type thermal conductivity,
Knudsen number and Wiedemann-Franz law. Finally, 
we conclude our work in section \ref{conclusion}.
\section{QUASIPARTICLE MODEL FOR HOT AND DENSE QCD MATTER}\label{quasiparticle}
At asymptotically high temperature, a system of 
quarks and gluons can be treated as an ideal gas 
due to asymptotic freedom. The interaction among 
quasi quarks and quasi gluons can be incorporated 
through medium dependent mass of quasiparticles 
which can be evaluated using one-loop perturbative 
thermal QCD. In pure thermal medium at finite 
quark chemical potential ($\mu$), the thermally
 generated mass for quarks and gluons obtained to be 
 as \cite{bellac}
\begin{align}
	& m_{th}^2 = \frac{1}{8}g^2C_F\Big(T^2 + \frac{\mu^2}{\pi^2}\Big),\nonumber\\
	& m_g^2 = \frac{1}{6}g^2T^2\left(C_A + \frac{1}{2}N_f\right),
\end{align}
respectively, where $C_F = \left(N_c^2 - 1\right)/2N_C = \frac{4}{3}$ for $N_C =3$, $C_A (C_A=3)$ is the group factor, $N_f$ is the 
number of flavor, $g$ is the QCD coupling constant 
with $g^2 = 4\pi\alpha_s$, where $\alpha_s$ is the
 one-loop running coupling constant, which runs with 
 temperature as \cite{a.ayala}
\begin{equation}
	\alpha_s(\Lambda^2) = \frac{1}{b_1\ln\Big(\frac{\Lambda^2}{\Lambda_{\overline{MS}}^2}\Big)},
\end{equation}
where $b_1 = \left(11 N_c - 2 N_f\right)/12\pi$ and 
$\Lambda_{\overline{MS}} =$ 0.176 GeV. The renormalization 
scale for quarks and gluons is chosen to 
be $\Lambda_q$ = $2\pi\sqrt{T^2 + \mu^2/\pi^2}$ 
and $\Lambda_g$ = $2\pi T$ respectively. Further, the dispersion relation of fermions in pure thermal medium (B=0) in the low ($p<<m_{th}$) momentum and high momentum ($p>>m_{th}$) limit are given as \cite{bellac,Blaziot} 
\begin{align}
	\omega_+(p) = m_{th} + \frac{p}{3}; \qquad  p<<m_{th}\\
	\omega_+(p) = p + \frac{m^2_{th}}{p}; \qquad p>>m_{th}.
\end{align}
As, we can see the thermal mass in both the low and high momentum limits is of the same order, $m_{th}\sim gT$.\\

\begin{figure}[h]\centering
	\includegraphics[width=0.35\textwidth]{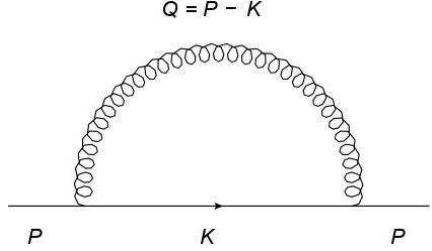}
	\caption{One-loop quark self energy in hot and magnetized medium}\label{self_quark}
\end{figure}

\nd The effective quark mass for $f$th 
 flavor can be written as \cite{bannur}
\begin{equation}
m^2_f = m_{f0}^2 + \sqrt{2}m_{f0}m_{f,th}+m_{f,th}^2,
\end{equation}
where $m_{f0}$ and $m_{f,th}$ is the current quark mass 
and thermal mass for $f$th flavor respectively. In 
presence of magnetic field, the one-loop running coupling constant, which runs with temperature and magnetic field, is given by \cite{a.ayala}
\begin{equation}
	\alpha_s(\Lambda^2,|eB|) = \frac{\alpha_s(\Lambda^2)}{1+b_1\alpha_s(\Lambda^2)\ln\left(\frac{\Lambda^2}{\Lambda^2+|eB|}\right)}.
\end{equation}
  The effective quark 
mass in presence of magnetic field can be generalized to
\begin{equation}\label{4}
	m^2_f = m_{f0}^2 + \sqrt{2}m_{f0}m_{fth,B}+m_{fth,B}^2
\end{equation}
 where $m_{fth,B}$ can be obtained by taking the static limit of denominator of the dressed quark propagator in magnetic 
 field. The inverse of the dressed quark propagator using Schwinger-Dyson equation can be written as
 \begin{align}\label{inverse_prop}
 		S^{*-1}(P) &= S^{-1}(P) - \Sigma(P)\nonumber\\
 	 &= \slashed{P} - \Sigma(P)
 \end{align}
where $S^{-1}(P)$ is bare inverse propagator and $\Sigma(P)$ is the quark self energy. So, to calculate the effective quark propagator in presence of magnetic field at finite temperature we need to evaluate the quark self energy as shown in Fig.\eqref{self_quark}.
\noindent The quark propagator in presence of background 
 magnetic field following the Schwinger formalism 
 can be written in terms of Laguerre 
 polynomial ($L_l(2\alpha)$) \cite{shwinger}
 \begin{equation}
 	iS(K) = \sum_{l=0}^{\infty} \frac{-id_l(\alpha) D + d^{\prime}_l(\alpha)\bar{D}}{k_L^2+2l|q_fB|}+\frac{i\gamma.k_{\perp}}{k_{\perp}^2},
 \end{equation}
where $q_f$ is the absolute charge of $f$th flavor, 
$l$ = 0, 1, 2, $\dots$ are the Landau levels, $||$ 
and $\perp$ are the parallel and perpendicular 
components of momentum respectively with respect 
to direction of magnetic field, $\alpha=k_{\perp}^2/|q_fB|$, $k_L^2 = m_{f0}^2-k_{||}^2$ and $d_l(\alpha), d^{\prime}_l(\alpha), D, \bar{D}$ 
are given as \cite{kang},

\begin{align}
	&d_l(\alpha) = (-1)^l e^{-\alpha}C_l(2\alpha),\nonumber\\
	&d^{\prime}_l(\alpha) = \frac{\partial d_l}{\partial\alpha},\nonumber\\
	& D = (m_{f0}+\gamma.k_{||})+ \gamma.k_{\perp}\left(\frac{m_{f0}^2-k_{||}^2}{k_{\perp}^2}\right),\nonumber\\
	&\bar{D} = \gamma_1\gamma_2\left(m_{f0} + \gamma.k_{||}\right),
\end{align}
with $C_l(2\alpha) = L_{l}(2\alpha) - L_{l-1}(2\alpha)$. In weak 
field limit, the quark propagator can be reorganized 
in power series of magnetic field $\left({\bf{q_fB}}\right)$ as,
\begin{equation}\label{fermion_prop_magnetic}
	iS(K) = \frac{i\left(\slashed{K}+m_{f0}\right)}{K^2-m_{f0}^2} - \frac{\gamma_1\gamma_2\left(\gamma.K_{||}+m_{f0}\right)}{\left(K^2-m_{f0}^2\right)^2}(q_fB),
\end{equation}
where first term in Eq.\eqref{fermion_prop_magnetic} is 
the free fermion propagator and second term is the $\mathcal{O}(q_fB)$ correction to it. Neglecting the current quark mass 
under the limit $\left(m_{f0}^2<q_fB<T^2\right)$ 
in the numerator and using the following metric tensor in Eq.\eqref{fermion_prop_magnetic},
\begin{align}
	g^{\mu\nu} = & g^{\mu\nu}_{||} + g^{\mu\nu}_{\bot};\nonumber\\
	g^{\mu\nu}_{\parallel} = \text{diag}(1,0,0,-1); &\quad g^{\mu\nu}_{\bot} = \text{diag}(0,-1,-1,0);\nonumber\\ 
	p^{\mu} = p^{\mu}_{\parallel} + p^{\mu}_{\bot}; &\quad p^{\mu}_{\parallel} = (p^0,0,0,p^3);\nonumber\\
	p^{\mu}_{\bot} = (0,p^1,p^2,0); &\quad \slashed{p} = \gamma^{\mu}p_{\mu}=\slashed{p}_{\parallel} + \slashed{p}_{\bot};\nonumber\\
	\slashed{p}_{\parallel} = \gamma^0 p_0 - \gamma^3 p^3; &\quad \slashed{p}_{\bot} = \gamma^1 p^1 + \gamma^2 p^2,
\end{align}
with
\begin{equation}
	i\gamma_1\gamma_2\slashed{K}_{||} = -\gamma_5[(K.b)\slashed{u}-(K.u)\slashed{b}],
\end{equation}
we obtain the quark propagator in presence of magnetic field at finite temperature as
\begin{equation}\label{quark_prop}
	iS(K) = \frac{i\slashed{K}}{K^2-m_{f0}^2} - \frac{i\gamma_5[(K.b)\slashed{u}-(K.u)\slashed{b}]}{(K^2-m_{f0}^2)^2}\big(q_f B\big),
\end{equation}
where $u^{\mu}= (1,0,0,0)$ denotes local rest frame of the heat bath. Introduction of a particular frame of reference breaks the Lorentz symmetry of the system. Similarly, $b^{\mu}=(0,0,0,1)$ denotes the preferred direction 
of magnetic field in our system which then breaks the rotational 
symmetry of the system. Using the quark propagator \eqref{quark_prop}, the one-loop quark self energy upto $\mathcal{O}(q_fB)$ 
in hot and weakly magnetized medium can be written as
\begin{align}\label{self_energy}
	\Sigma(P) =  g^2 C_F T\sum_n\int\frac{d^3k}{(2\pi)^3}\gamma_{\mu}\Bigg(\frac{\slashed{K}}{(K^2-m_{f0}^2)}
	-& \frac{\gamma_5[(K.b)\slashed{u}-(K.u)\slashed{b}]}{(K^2-m_{f0}^2)^2}(q_fB)\Bigg)\times&\nonumber\\
	\gamma^{\mu}\frac{1}{(P-K)^2},
\end{align}
where $T$ is the temperature of the system and $g^2 = 4\pi\alpha_s(\Lambda^2,|eB|)$. 
First term is the thermal medium contribution ($\Sigma_{0}$)  
whereas second one is with magnetic field correction term ($\Sigma_{1}$).\par
 The general covariant structure of quark self energy 
 at finite temperature and magnetic field can 
 be written as \cite{aritra}
\begin{equation}\label{52}
	\Sigma(P) = -\mathcal{A}\slashed{P}-\mathcal{B}\slashed{u}-\mathcal{C}\gamma_5\slashed{u}-\mathcal{D}\gamma_5\slashed{b},
\end{equation} 
where $\mathcal{A}, \mathcal{B}, \mathcal{C}, \mathcal{D}$ 
are the structure functions. Using Eq.\eqref{self_energy} 
and \eqref{52}, the general form of these structure 
functions are obtained as 
\begin{align}\label{A}
	& \mathcal{A}\left(p_0, p_{\perp}, p_z\right)= \frac{1}{4}\frac{\text{Tr}(\Sigma(P)\slashed{P})-(P.u)\text{Tr}(\Sigma(P)\slashed{u})}{(P.u)^2-P^2},\\ \label{B}
	& \mathcal{B}\left(p_0, p_{\perp}, p_z\right)= \frac{1}{4}\frac{\left(-P.u\right)\text{Tr}(\Sigma(P)\slashed{P})+P^2\text{Tr}(\Sigma(P)\slashed{u})}{(P.u)^2-P^2},\\ \label{C}
	& \mathcal{C}\left(p_0, p_{\perp}, p_z\right)= -\frac{1}{4}\text{Tr}(\gamma_5\Sigma(P)\slashed{u}),\\ \label{D}
	& \mathcal{D}\left(p_0, p_{\perp}, p_z\right)= \frac{1}{4}\text{Tr}(\gamma_5\Sigma(P)\slashed{b}).
\end{align}
These structure functions are found to depend upon 
various Lorentz scalars defined by
\begin{align}
	& p^0\equiv P^{\mu}u_{\mu}=\omega,\\
	& p^3\equiv P^{\mu}b_{\mu}= -p_z,\\
	& p_{\bot}\equiv\big[(P^{\mu}u_{\mu})^2 - (P^{\mu}b_{\mu})^2 - (P^{\mu}P_{\mu})^2\big]^{1/2},
\end{align}
where $\omega,p_{\bot},p_{z}$ are termed as Lorentz 
invariant energy, transverse momentum and longitudinal 
momentum respectively. The detailed calculation of 
all these structure functions is shown in 
Appendix \ref{append} and results are quoted here,
\begin{align}\label{A_net}
	& \mathcal{A}(p_0,|{\bf{p}}|) = \frac{m_{th}^2}{|{\bf{p}}|^2}Q_1\Bigg(\frac{p_0}{|{\bf{p}}|}\Bigg),\\
	& \mathcal{B}(p_0,|{\bf{p}}|) =- \frac{m_{th}^2}{|{\bf{p}}|}\Bigg[\frac{p_0}{|{\bf{p}}|}Q_1\Big(\frac{p_0}{|{\bf{p}}|}\Big)-Q_0\Bigg(\frac{p_0}{|{\bf{p}}|}\Bigg)\Bigg],\\
	& \mathcal{C}(p_0,|{\bf{p}}|) = -4g^2C_FM^2\frac{p_z}{|{\bf{p}}|^2}Q_1\Bigg(\frac{p_0}{|{\bf{p}}|}\Bigg),\\
	& \mathcal{D}(p_0,|{\bf{p}}|) = -4g^2C_FM^2\frac{1}{|{\bf{p}}|}Q_0\Bigg(\frac{p_0}{|{\bf{p}}|}\Bigg),\label{D_net}
\end{align}
where $Q_0$ and $Q_1$ are Legendre functions of first
 and second kind respectively read as
\begin{align}
	&Q_0(x) = \frac{1}{2}\ln\left(\frac{x+1}{x-1}\right),\\
	&Q_1(x) = \frac{x}{2}\ln\left(\frac{x+1}{x-1}\right)-1 = xQ_0(x)-1,
\end{align}
with magnetic mass obtained as \cite{aritra_bandho} 
\begin{align}
	& M^2(T,\mu,m_{f0},q_fB) = \frac{|q_fB|}{16\pi^2}\left(\frac{\pi T}{2m_{f0}} -\text{ln} 2 + \frac{7\mu^2\zeta(3)}{8\pi^2T^2}\right).
\end{align}
where $\zeta$ is the Riemann zeta function. 
The general 
covariant structure of quark self energy Eq.(\ref{52})
 can be recast in terms of 
 left handed ($P_L =(\mathbb{I}-\gamma_5)/2$) 
 and right handed ($P_R =(\mathbb{I}+\gamma_5)/2$) 
 chiral projection operators as
\begin{equation}\label{63}
	\Sigma(P) = -P_R\slashed{A'}P_L - P_L\slashed{B'}P_R,
\end{equation}
with $\slashed{A'}$ and $\slashed{B'}$ defined as
\begin{align}
	& \slashed{A'} = \mathcal{A}\slashed{P} + (\mathcal{B}+\mathcal{C})\slashed{u} + \mathcal{D}\slashed{b},\\
	& \slashed{B'} = \mathcal{A}\slashed{P} + (\mathcal{B}-\mathcal{C})\slashed{u} - \mathcal{D}\slashed{b}.
\end{align}
Using Eq.\eqref{inverse_prop} and (\ref{63}), inverse fermion propagator can be written as
\begin{align}
	S^{*-1}(P) = \slashed{P} + P_R\left[\mathcal{A}\slashed{P} + \left(\mathcal{B}+\mathcal{C}\right)\slashed{u} + \mathcal{D}\slashed{b}\right]P_L + P_L\left[\mathcal{A}\slashed{P} + \left(\mathcal{B}-\mathcal{C}\right)\slashed{u} - \mathcal{D}\slashed{b}\right]P_R,
\end{align}
and using $P_{L,R}\gamma^{\mu} = \gamma^{\mu} P_{R,L}$ and $P_L \slashed{P} P_L = P_R \slashed{P} P_R = P_L P_R \slashed{P} = 0$, 
we obtain
\begin{equation}
		S^{*-1}(P) = P_R \slashed{L} P_L + P_L \slashed{R} P_R,
\end{equation}
where $\slashed{L}$ and $\slashed{R}$ are 
\begin{align}
	&\slashed{L} = (1+\mathcal{A})\slashed{P} + (\mathcal{B}+\mathcal{C})\slashed{u} + \mathcal{D}\slashed{b},\\
	& \slashed{R} = (1+\mathcal{A})\slashed{P} + (\mathcal{B}-\mathcal{C})\slashed{u} - \mathcal{D}\slashed{b}.
\end{align}
Thus, we get the effective quark propagator as
\begin{equation}
	S^*(P) = \frac{1}{2}\left[ P_L\frac{\slashed{L}}{L^2/2}P_R+P_R\frac{\slashed{R}}{R^2/2}P_L\right],
\end{equation}
where 
\begin{align}
	& L^2 = (1+\mathcal{A})^2P^2 + 2(1+\mathcal{A})(\mathcal{B}+\mathcal{C})p_0-2\mathcal{D}(1+\mathcal{A})p_z+ (\mathcal{B}+\mathcal{C})^2-\mathcal{D}^2,\\
	& R^2 = (1+\mathcal{A})^2P^2 + 2(1+\mathcal{A})(\mathcal{B}-\mathcal{C})p_0+2\mathcal{D}(1+\mathcal{A})p_z+ (\mathcal{B}-\mathcal{C})^2-\mathcal{D}^2.
\end{align}
Next, we take the static limit 
($p_0 =0,|{\bf{p}}|\rightarrow 0$) of $L^2/2$ and $R^2/2$,
 after expanding the Legendre functions involved 
 in structure functions in power series of 
 $\frac{|\bf{p}|}{p_0}$. Considering upto $\mathcal{O}(g^2)$, 
 we obtain
\begin{align}
	&\frac{L^2}{2}\arrowvert_{p_0= 0, {|\bf{p}|}\rightarrow 0} = m_{th}^2 + 4g^2 C_F M^2,\\
	&\frac{R^2}{2}|_{p_0= 0, {|\bf{p}|}\rightarrow 0} = m_{th}^2 - 4g^2 C_F M^2.
\end{align}
The degenerate left- and right- handed modes get 
separated out in presence of weak magnetic field
 and hence the thermal mass (squared) at finite 
 chemical potential in presence of weak magnetic 
 field obtained as
\begin{align}\label{qaurk_mass}
	&m_{L}^2 = m_{th}^2 + 4g^2 C_F M^2,\\
	&m_{R}^2 = m_{th}^2 - 4g^2 C_F M^2,
\end{align}
which is opposite to the case of strong magnetic 
field where left- and right-handed chiral modes 
have the same mass \cite{S_rath}. The quasiparticle mass obtained above consists of pure thermal and magnetic contributions. The thermal mass is independent of chiral modes whereas magnetic contribution depends on the chiral modes. The dispersion relation for both chiral modes of quarks at low and high momentum limit in lowest Landau level is given by \cite{aritra}
\begin{align}
	\omega_{L/R}(p_z) = m_{L/R} + \frac{p_z}{3} ; \qquad (p_z<<m_{L/R})\\
	\omega_{L/R}(p_z) = |p_z| + \frac{m_{L/R}^2}{p_z}; \qquad (p_z>>m_{L/R}).
\end{align}
 The form of dispersion relations in both low and high momentum limits even in the presence of magnetic field is similar to that in the absence of magnetic field and the masses in both the limits are of the same order. However, the quasiparticle mass used in our computation is obtained from the static limit of the pole of the full quark propagator (up to one-loop quark self energy), which is independent of low momentum and high momentum limits. Now, we will 
evaluate the charge and thermal transport 
coefficients in the presence of weak magnetic 
field at finite chemical potential for left- 
and right-handed modes separately as the system will not be in both modes simultaneously.

\section{CHARGE TRANSPORT COEFFICIENTS}\label{charge_transport}
Boltzmann transport equation governs the evolution 
of phase space density $f(x,p)$ associated 
with the partons in our system. The QGP is a relativistic plasma, since $T>>$ any mass scale in the system. This validates the use of the relativistic Boltzmann transport equation (RBTE) to carry out our investigation. The RBTE 
for relativistic particle with charge $q$ in 
presence of external electromagnetic field can 
be written as
\cite{landau}
\begin{equation}\label{1}
p^{\mu}\partial_{\mu}f(x,p) + q F^{\mu\nu}p_{\nu}\frac{\partial f(x,p)}{\partial p^{\mu}} = C[f],
\end{equation}

where, $f$ is 
the distribution function deviated slightly 
from equilibrium distribution function ($f_0$) with $f = f_0 + \delta f$ ($\delta f<<f_0$).
$F^{\mu\nu}$ is the anti-symmetric electromagnetic 
field tensor, $C[f]$ is the collision integral 
that describes the rate of change of distribution 
function by virtue of collisions. The general form of 
collision integral consists of absorption and emission 
terms in phase space volume element. This leads to
 nonlinear integro-differential equation which is 
 very complicated to solve. Hence, we simplify the 
 equation using relaxation-time approximation (RTA). 
 Under relaxation time approximation, the external
  perturbation takes the system slightly away 
  from equilibrium from which it relaxes towards 
  the equilibrium exponentially with time scale $\tau$. 
  Collision integral under RTA takes the form as
\begin{equation}\label{2}
C[f]\simeq - \frac{p^{\mu}u_{\mu}}{\tau} (f-f_0)\equiv -\frac{p^{\mu}u_{\mu}}{\tau}\delta f,
\end{equation}
 where $u_{\mu}$ is the fluid 4-velocity, $\tau$ is the thermal averaged 
 relaxation time.  The Eq.(\ref{1}) in 3-notation can 
 be written as

\begin{equation}\label{3}
\frac{\partial f}{\partial t} + {\bf{v}}.\frac{\partial f}{\partial \bf{r}}  + {q(\bf{E + v\times B})}.\frac{\partial f}{\partial \bf{p}} = -\frac{1}{\tau} (f-f_0).
\end{equation}

We consider the spatially uniform $\frac{\partial f}{\partial{\bf{r}}}\approx 0$ and static medium $\frac{\partial f}{\partial t} = 0$ 
such that there are no space-time gradient. 
Then Eq.(\ref{3}) simplifies to
\begin{equation}
q({\bf{E + v\times B}}).\frac{\partial f}{\partial{\bf{p}}} = -\frac{1}{\tau}(f-f_0).
\end{equation}
For the sake of simplicity of calculation, we choose the transverse electric 
 and magnetic field as ${\bf{E}} = E\hat{x}$ and ${\bf{B}} = B\hat{z}$. 
 This yields:
\begin{equation}\label{5}
f-qB\tau\Big(v_x\frac{\partial f}{\partial p_y} - v_y\frac{\partial f}{\partial p_x}\Big) = f_0 - qE\tau\frac{\partial f_0}{\partial p_x}.
\end{equation}
In order to solve Eq.(\ref{5}), we take the following 
\textit{ansatz} of distribution function $f(p)$ as \cite{feng}
\begin{equation}\label{6}
f(p) = f_0 - \tau q{\bf{E}}.\frac{\partial f_0}{\partial{\bf{p}}} - {\bm{\xi}}.\frac{\partial f_0}{\partial{\bf{p}}},
\end{equation}
and $f_{0}$ is given by,
\begin{equation}
f_0({\bf{p}}) = \frac{1}{e^{(\sqrt{{\bf{p^2}}+m^2}-\mu)/T}+1},
\end{equation}
which is space and time independent solution to the 
Boltzmann equation and $f_0$ satisfies,
\begin{equation}
\frac{\partial f_0}{\partial\bf{p}} = {\bf{v}}\frac{\partial f_0}{\partial\varepsilon}  , \quad \frac{\partial f_0}{\partial\varepsilon} = -\beta f_0(1-f_0),
\end{equation}
where $\varepsilon =\sqrt{{\bf{p^2}} + m^2}$ is the single 
particle energy. Using the \textit{ansatz} Eq.\eqref{6} in Eq.(\ref{5}), 
we get
\begin{align}\label{50}
\left(f_0 - \tau q{\bf{E}}.\frac{\partial f_0}{\partial{\bf{p}}} - \bm{\xi}.\frac{\partial f_0}{\partial\bf{p}}\right) - qB\tau\left(v_x\frac{\partial}{\partial p_y} - v_y\frac{\partial}{\partial p_x}\right)&\left(f_0 - \tau q{\bf{E}}.\frac{\partial f_0}{\partial{\bf{p}}} - \bm{\xi}.\frac{\partial f_0}{\partial\bf{p}}\right)\nonumber\\
&= f_0 - qE\tau\frac{\partial f_0}{\partial p_x}.
\end{align}
The first term in the parenthesis in left hand side of 
Eq.\eqref{50} can be rewritten as,
\begin{equation}\label{51}
\left(f_0 - \tau q{\bf{E}}.\frac{\partial f_0}{\partial{\bf{p}}} - \bm{\xi}.\frac{\partial f_0}{\partial\bf{p}}\right)=	f_0 + \beta\tau qE v_x f_0 + (\bm{\xi} .{\bf{v}})\beta f_0.
\end{equation}  
Neglecting the $f_0^2$ terms at high temperature and 
using the following second order partial derivatives, 
\begin{align}
	&\frac{\partial^2 f_0}{\partial p_y p_x} = \frac{\beta p_x p_y f_0}{\varepsilon^2}\left(\beta + \frac{1}{\varepsilon}\right),\\
	&\frac{\partial^2 f_0}{\partial p_y p_z} = \frac{\beta p_y p_z f_0}{\varepsilon^2}\left(\beta + \frac{1}{\varepsilon}\right),\\
	&\frac{\partial^2 f_0}{\partial p_y^2} = -\beta\left[\frac{f_0}{\varepsilon}-\frac{f_0 p_y^2}{\varepsilon^2}\left(\beta + \frac{1}{\epsilon}\right)\right],
\end{align}
the second term in left hand side of Eq.\eqref{50} 
get reduced to 
\begin{equation}\label{55}
	qB\tau\left(v_x\frac{\partial}{\partial p_y} - v_y\frac{\partial}{\partial p_x}\right)\left(f_0 - \tau q{\bf{E}}.\frac{\partial f_0}{\partial{\bf{p}}} - \bm{\xi}.\frac{\partial f_0}{\partial\bf{p}}\right) = qB\tau\beta\left(\frac{\xi_y v_x }{\varepsilon}- \frac{\xi_x v_y}{\varepsilon}-\frac{ v_y\tau qE }{\varepsilon}\right) f_0.
\end{equation}
Combining the Eq.\eqref{51} and \eqref{55}, Eq.\eqref{50} 
obtained as
\begin{equation}\label{9}
\frac{\tau qB qE v_y}{\varepsilon} - \frac{qB}{\varepsilon}(v_x\xi_y - v_y\xi_x) + \frac{1}{\tau}\Big(\xi_x\frac{p_x}{\varepsilon} + \xi_y\frac{p_y}{\varepsilon} + \xi_z\frac{p_z}{\varepsilon}\Big) = 0.
\end{equation}
 The above equation should be satisfied for any value 
 of velocity therefore, comparing the coefficients of
  $v_x, v_y$ and $v_z$ of Eq.\eqref{9}, we get 
\begin{align}
&\xi_z = 0,\\
&\omega_c \tau qE + \omega_c \xi_x + \frac{\xi_y}{\tau} = 0,\\
&\frac{\xi_x}{\tau} - \omega_c\xi_y = 0,
\end{align}
where $\omega_c = \frac{qB}{\varepsilon}$ is termed as 
cyclotron frequency. Solving for $\xi_x$ and $\xi_y$, we have
\begin{equation}\label{12}
\xi_x = \frac{-\omega_c^2\tau^3qE}{(\omega_c^2\tau^2 +1)}; \quad \xi_y = \frac{-\omega_c\tau^2 qE}{(\omega_c^2\tau^2 +1)}.
\end{equation}
Using Eq.(\ref{12}) in Eq.(\ref{6}), the distribution 
function $f(p)$ for quarks simplifies to
\begin{equation}\label{13}
f(p) = f_0 - \frac{qEv_x\tau}{(1+ \omega_c^2\tau^2)}\left(\frac{\partial f_0}{\partial\varepsilon}\right) + \frac{qEv_y\omega_c\tau^2}{(1 +\omega_c^2\tau^2)}\left(\frac{\partial f_0}{\partial\varepsilon}\right),
\end{equation}
and for anti-quarks ($f\rightarrow\bar{f}, q\rightarrow -q, \omega_c\rightarrow -\omega_c$),
\begin{equation}
	\bar{f}(p) = \bar{f_0} + \frac{qEv_x\tau}{(1+ \omega_c^2\tau^2)}\left(\frac{\partial \bar{f_0}}{\partial\varepsilon}\right) + \frac{qEv_y\omega_c\tau^2}{(1 +\omega_c^2\tau^2)}\left(\frac{\partial \bar{f_0}}{\partial\varepsilon}\right).
\end{equation}
The induced current in the system as a result of external 
electromagnetic fields can be written as
\begin{align}\label{14}
j^i  = \sigma_{\text{Ohmic}}\delta^{ij}E_j + \sigma_{\text{Hall}}\epsilon^{ij}E_{j},
\end{align}
where $\sigma_{\text{Ohmic}}$ and $\sigma_{\text{Hall}}$ 
are Ohmic and Hall conductivities respectively, 
$\epsilon_{ij}$ is the $2\times2$ antisymmetric 
unity tensor, with $\epsilon_{12}=-\epsilon_{21}=1$. 
It is clear from the above equation that 
$\sigma_{{\text{Ohmic}}}$ describes the longitudinal 
response (current along the direction of electric field) and $\sigma_{{\text{Hall}}}$ describes the transverse 
response (current transverse to the electric field). 
Further, the induced current can be written in 
terms of deviation $\delta f$ ($\delta \bar{f}$) 
from $f_0$ ($\bar{f_0}$) as
\begin{equation}\label{64}
{\bf{j}} = g_f\int\frac{d^3{p}}{(2\pi)^3}{\bf{v}} \left(q\delta f(p) + \bar{q}\delta\bar{f}(p)\right),
\end{equation}
where $\bar{q}$ is the charge of antiparticle and $g_f=3\times 2$ is the color and spin degeneracy 
factor of fermions. Using Eq.(\ref{14}) and (\ref{64}), the 
Ohmic and Hall conductivity for a system of multiple 
charge species can be written as
\begin{equation}\label{15}
\sigma_{{\text{Ohmic}}} = \frac{1}{6\pi^2T}\sum_fg_fq^2_{f}\tau_f\int dp\frac{p^4}{\varepsilon_f^{2}}\frac{1}{(1 +\omega_c^2\tau_f^2)}[f^0_f(1-f^0_f)+\bar{f^0_f}(1-\bar{f^0_f})],
\end{equation}
\begin{equation}\label{16}
\sigma_{{\text{Hall}}} = \frac{1}{6\pi^2T}\sum_fg_fq^2_{f}\tau_f^2\int dp\frac{p^4}{\varepsilon_f^{2}}\frac{\omega_c}{(1 +\omega_c^2\tau_f^2)}[f^0_f(1-f^0_f)-\bar{f^0_f}(1-\bar{f^0_f})],
\end{equation}
where $f$ stands for flavor and here we have 
used $f=$ up ($u$), down ($d$). The relaxation time for quarks (antiquarks) used above for calculation of conductivities is given by \cite{kajantie}, where massless $u$- and $d$-quarks were considered with $\mu<<T$,
\begin{equation*}
	\tau_{q(\bar{q})} = \frac{1}{5.1 T\alpha_s^2\log(\frac{1}{\alpha_s})(1+0.12(2N_f+1))}.
\end{equation*}
 It was argued in \cite{berreharah} that finite parton mass has little effect on scattering cross-section and hence on relaxation time. This leads to qualitatively same result for massless and massive partons. The current light quark ($m_{u,d}$) masses are chosen to be $0.1$ times the strange quark mass ($m_{s0}$) which is in compliant with chiral perturbation theory \cite{cheng,schmidt}. The parameters were adjusted to get the best fitted lattice data with $m_{s0}= 80$MeV \cite{zhu}. The Ohmic and 
Hall conductivity obtained above is defined 
by current which appears due to the effect 
of an electric and magnetic field when 
there is no temperature gradient or we can 
say isothermal Ohmic and Hall conductivity. 
As discussed in quasiparticle model, we will 
incorporate the quasiparticle mass which were 
obtained to be different for left- and right-handed 
chiral modes. $u(\bar{u})$ and $d(\bar{d})$ 
quarks are spin-$\frac{1}{2}$ particles and 
they can assume right-handedness or left-handedness 
depending on their up or down spin with respect 
to their direction of motion. We have taken into 
account the both modes for up and down quarks and 
computed the conductivities for L- and R-mode 
separately. In case of baryonic symmetry i.e. at 
zero quark chemical
potential, the distribution function for quarks 
and anti-quarks become equal and hence
Hall conductivity vanishes.
\section{HEAT TRANSPORT COEFFICIENTS}\label{thermal_transport}
In nonrelativistic case, the heat equation 
is obtained by the validity of the first and 
second laws of thermodynamics, where the flow 
of heat is proportional to the temperature 
gradient and the proportionality factor is 
called the thermal conductivity. We are intended to study the thermal conductivity in the system of partons.
The heat flow 4-vector which is defined to be 
difference between energy diffusion and 
enthalpy diffusion is given as \cite{s_groot},
\begin{equation}
Q_{\mu} = \Delta_{\mu\alpha}T^{\alpha\beta}u_{\beta}-h\Delta_{\mu\alpha}N^{\alpha},
\end{equation}
where $\Delta_{\mu\alpha} = g_{\mu\alpha}-u_{\mu}u_{\alpha}$ 
is the projection operator. $N^{\alpha}$ is characterized as first moment of distribution function which corresponds to the particle four-flow vector as
\begin{equation}\label{18}
	N^{\alpha} = \sum_f g_f \int\frac{d^3 p}{(2\pi)^3}\frac{p^{\alpha}}{\varepsilon_f}\left[f_f - \bar{f}_f\right],
\end{equation}
whereas, $T^{\alpha\beta}$ is characterized as second moment of distribution function which corresponds to energy-momentum tensor as
 \begin{equation}\label{19}
 	T^{\alpha\beta}=\sum_f g_f\int\frac{d^3p}{(2\pi)^3}\frac{p^{\alpha}p^{\beta}}{\varepsilon_f}\left(f_f +\bar{f}_f\right)+g_g\int\frac{d^3p}{(2\pi)^3}\frac{p^{\alpha}p^{\beta}}{\varepsilon_g}f_g,
 \end{equation}
where, $f_f, \bar{f}_f$ and $f_g$ are quark, antiquark and gluon distribution function, $\varepsilon_g$ is the single particle energy for gluons and $g_g = 8\times 2$ is the degeneracy factor of the gluons. We can obtain the particle number density, energy density 
and pressure from the above equation as $n=N^{\alpha}u_{\alpha}$, $e=u_{\alpha}T^{\alpha\beta}u_{\beta}$ and $P=-\Delta_{\alpha\beta}T^{\alpha\beta}/3$ respectively. Therefore, enthalpy per particle can be obtained as, $h=(e+P)/n$.
The heat flow 4-vector in rest frame of heat 
bath is orthogonal to fluid 4-velocity, i.e. 
$Q_{\mu}u^{\mu} =0$. Thus, heat flow is spatial 
which can be written in terms of infinitesimal
 changes in the distribution function as
\begin{align}\label{20}
{\bf{Q}} = \sum_f g_f \int\frac{d^3p}{(2\pi)^3}\frac{{\bf{p}}}{\varepsilon_f}\left[\left(\varepsilon_f - h_f\right)\delta f_f + \left(\varepsilon_f + \bar{h}_f\right)\delta\bar{f}_f\right]+g_g\int\frac{d^3p}{(2\pi)^3}{\bf{p}}\delta f_g.
\end{align}
 The particle four-flow and energy-momentum
tensor can be decomposed with respect to an arbitrary
normalized time-like four vector, $u^{\mu} = \gamma (1,\vec{v})$, where $u^{\mu}u_{\mu} =1$. The most general decomposition is given as \cite{Landau_fluid,Greif}
\begin{align}
	&N^{\mu} = n u^{\mu} + v^{\mu},\\
	&T^{\mu\nu} = -Pg^{\mu\nu} + w u^{\mu}u^{\nu} + \pi^{\mu\nu},
\end{align}
where, $v^{\mu}$ is the particle diffusion current, $w = e+P$ 
is the heat function per unit volume, $\pi^{\mu\nu}$ is the 
shear-stress tensor. Using the continuity equation and law 
of increase of entropy, the required form of symmetrical 
4-tensor $\pi^{\mu\nu}$ and 4-vector $v^{\mu}$ obtained 
to be as \cite{Landau_fluid},
\begin{align}
	&\pi^{\mu\nu} = -c\eta\left(\frac{\partial u^{\mu}}{\partial x_{\nu}} + \frac{\partial u^{\nu}}{\partial x_{\mu}} - u^{\mu}u_{\alpha}\frac{\partial u^{\nu}}{\partial x_{\alpha}} - u^{\mu}u^{\alpha}\frac{\partial u^{\mu}}{\partial x^{\alpha}}\right) - \left(\zeta-\frac{2}{3}\eta\right)\frac{\partial u^{\alpha}}{\partial x^{\alpha}}\left(g^{\mu\nu}-u^{\mu}u^{\nu}\right),\\
	&v^{\mu} = \frac{\kappa}{c}\left(\frac{nT}{w}\right)^2\left[\frac{\partial}{\partial x_{\mu}}\left(\frac{\mu}{T}\right)-u^{\mu}u^{\alpha}\frac{\partial}{\partial x^{\alpha}}\left(\frac{\mu}{T}\right)\right].\label{12}
\end{align}
Here, $\eta$ and $\zeta$ are shear and bulk viscosity 
coefficients and $\kappa$ is the thermal conductivity, 
taken in accordance with their non-relativistic definitions.
The zero particle flux ($nu^i + v^i =0$) corresponds to the pure thermal conduction. The spatial components $u^{i} = -v^{i}/n$ 
of the 4-velocity are of the first order in the gradients; 
since $\pi^{\mu\nu}, v^{\mu}$ are written only as far 
as this order, the 4-velocity component $u^0$ must be
taken as unity. To the same accuracy, omitting the 
second term in the square bracket of Eq.\eqref{12},
the energy flux density from $T^{\mu\nu}$ is given as,
\begin{align}
	&cT^{0i} = cwu^0u^i =- cwv^i/n\\\nonumber
	&=- \kappa \frac{nT^2}{e+P}\left[\frac{\partial}{\partial x_i}\left(\frac{\mu}{T}\right)\right],
\end{align}
which relates the energy flux density/heat flow with 
the gradient of thermodynamical potential ($U = \mu/T$) as in Navier-Stokes theory. In terms of 4-gradient, the above equation can be written as 
\begin{align}\label{71}
Q_{\mu}=&-\kappa\frac{nT^2}{e + P}\nabla_{\mu}U,
\end{align}
 where $\kappa$ is the thermal conductivity and 
 $\nabla_{\mu}$ is the 4-gradient, $\nabla_{\mu} = \partial_{\mu}-u_{\mu}u_{\nu}\partial^{\nu}$.
  The entropy density ($s = s(e, n)$) 
  in equilibrium state in terms of energy density,
   pressure and chemical potential can be written
    as \cite{israel},
 \begin{equation}\label{72}
 	s = \left(\frac{e + P}{T}\right) - \left(\frac{\mu}{T}\right) n.
 \end{equation} 
Further, the inverse of temperature ($T^{-1}$) and 
thermodynamic potential ($U$) can be defined as
 partial derivative of $s(e, n)$
\begin{equation}\label{73}
	\text{d}s = \frac{1}{T}\text{d}e - U \text{d}n .
\end{equation}
Using Eq.\eqref{72} and \eqref{73}, we obtain
\begin{align}
	 &\text{d}\left(\frac{P}{T}\right)= -e \text{d}\left(\frac{1}{T}\right) + n\text{d}U ,\nonumber\\
	 &\frac{\text{d}P}{nT} = \text{d}U + \frac{1}{T^2}\left(\frac{P + e}{n}\right)\text{d}T.
\end{align}
Generalizing it to the 4-gradient, the heat flow 
can be rewritten as
\begin{equation}
Q_{\mu}=\kappa\left[\nabla_{\mu}T-\frac{T}{e +P}\nabla_{\mu}P\right],
\end{equation}	
and in local rest frame, the spatial component of 
heat flow can be written as
\begin{equation}\label{22}
{\bf{Q}} = -\kappa\left[\frac{\partial T}{\partial{\bf{x}}} - \frac{T}{nh}\frac{\partial P}{\partial{\bf{x}}}\right].
\end{equation}
One can thus obtain the thermal conductivity ($\kappa$) 
by comparing Eq.(\ref{20}) and Eq.(\ref{22}). We will firstly calculate the contribution of quarks and anti-quarks to the thermal conductivity.
So now, expressing the relativistic Boltzmann transport equation in terms of gradients of flow velocity and temperature in
relaxation time approximation as
\begin{align}\label{23}
p^{\mu}\partial_{\mu}T\left(\frac{\partial f}{\partial T}\right) + 
& p^{\mu}\partial_{\mu} (p^{\nu}u_{\nu})\left(\frac{\partial f}{\partial p^0}\right)
 +q\Bigg(F^{0j}p_{j}\frac{\partial f}{\partial p^0}+ F^{j0}p_{0}\frac{\partial f}{\partial p^{j}} + \nonumber \\
& F^{ij}p_{j}\frac{\partial f}{\partial p^{i}}+F^{ji}p_{i}\frac{\partial f}{\partial p^{j}}\Bigg)= -\frac{p^{\mu}u_{\mu}}{\tau}\delta f,
\end{align}
where $p_0 = \varepsilon - \mu$ and for very small 
$\mu$, it can be approximated as $p_0\approx\varepsilon$. 
Using the following partial derivatives
\begin{equation}
\frac{\partial f_0}{\partial T} = \frac{\varepsilon}{T^2}f_0(1-f_0),
\end{equation}
\begin{equation}
\frac{\partial f_0}{\partial p^0} = -\frac{1}{T}f_0(1-f_0),
\end{equation}
\begin{equation}
\frac{\partial f_0}{\partial p_j} = -\frac{p_j}{Tp_0}f_0(1-f_0),
\end{equation}
the  Eq.(\ref{23}) is given as
\begin{align}\label{81}
-\frac{\delta f}{\tau} =&\frac{f_0(1-f_0)}{p^0}\left[p^{\mu}\partial_{\mu} T \left(\frac{p^0}{T^2}\right) - \frac{p^{\mu}\partial_{\mu}(p^{\nu}u_{\nu})}{T} - \frac{q}{T}\left(F^{0j}p_j + F^{j0} p_j\right)\right]+
\frac{q}{p^0}\left(F^{ij}p_{j}\frac{\partial f}{\partial p^{i}}+F^{ji}p_{i}\frac{\partial f}{\partial p^{j}}\right)\nonumber\\
&=\frac{1}{T}f_0(1-f_0)\Bigg[\frac{1}{T}\left(p^0\partial_0 T + p^j\partial_j T\right) + T\partial_{0}\left(\frac{\mu}{T}\right) +\frac{p^j T}{p^0}\partial_j\left(\frac{\mu}{T}\right)-\frac{1}{p^0}\left(p^0p^{\nu}\partial_0 u_{\nu}+ 
p^jp^{\nu}\partial_j u_{\nu}\right)-&\nonumber\\
&2q\frac{\textbf{E.p}}{p^0}\Bigg]
 + q\left(\textbf{v}\mathbb{\times}\textbf{B}\right).\frac{\partial f}{\partial\textbf{p}},
\end{align}
where $2F_{ij} = \epsilon_{ijk}B^k$. Now, exerting 
the energy-momentum conservation $\left(\partial_0 u_{\mu}=\frac{\nabla_{\mu}P}{nh}\right)$ along with 
relativistic Gibbs-Duhem relation,
\begin{equation*}
	\partial_j\left(\frac{\mu}{T}\right) = \frac{-h}{T^2}\left(\partial_j T- \frac{T}{nh}\partial_j P\right),
\end{equation*}
we obtain Eq.\eqref{81} as
\begin{align}\label{82}
-\frac{\delta f}{\tau}	=\frac{1}{T}f_0(1-f_0)\Bigg[\frac{1}{T}&\left(p^0\partial_0 T\right) + \left(\frac{p^0-h}{p^0}\right)\frac{p^j}{T}\left(\partial_j T-\frac{T}{nh}\partial_j P\right)+T\partial_0\left(\frac{\mu}{T}\right)-p^jp^{\nu}\left(\partial_j u_{\nu}\right)-2q\frac{\textbf{E.p}}{p^0}\Bigg]\nonumber\\
	&+ q\left(\textbf{v}\mathbb{\times}\textbf{B}\right).\frac{\partial f}{\partial\textbf{p}},
\end{align}
where $\frac{\partial f}{\partial\textbf{p}}=\frac{\partial}{\partial\textbf{p}}\left(f_0+\delta f\right)$ and $\frac{\partial f_0}{\partial p^j}\propto v^j$, 
therefore the Lorentz term vanishes for the 
equilibrium distribution function and we get,
\begin{align}\label{27}
-\frac{\delta f}{\tau} = &\frac{1}{T}f_0(1-f_0)\Bigg[\frac{p_0}{T}\partial_0 T + \left(\frac{p^0 - h}{p^0}\right)\frac{p^j}{T}\left(\partial_j T - \frac{T}{nh}\partial_j P\right)+ T\partial_0\Big(\frac{\mu}{T}\Big)\nonumber\\
& -\frac{p^jp^{\nu}}{p_0}(\partial_ju_{\nu})- \frac{2q}{p_0}({\bf{E.p}})\Bigg] + q({\bf{v\times B}})\frac{\partial(\delta f)}{\partial{\bf{p}}}.
\end{align}
Now, Choosing the \textit{ansatz} for infinitesimal deviation $(\delta f)$ from $f_0$ as \cite{m.kurian},
\begin{equation}\label{28}
\delta f = ({\bf{p}}.\bm{\chi})\frac{\partial f_0}{\partial\varepsilon},
\end{equation}
where $\bm{\chi}$ in turn is related to thermal driving 
force and magnetic field in medium and takes the form as
\begin{equation}\label{29}
\bm{\chi} = a_1{\bf{c}} + a_2{\bf{Y}} + a_3({\bf{Y}}\times{\bf{c}}).
\end{equation}
Here, ${\bf{c=\frac{B}{|B|}}}$ and ${\bf{Y}} = \frac{\bm{\nabla}T}{T}-\frac{\bm{\nabla}P}{nh}$. Using 
Eq.\eqref{28} and \eqref{27}, we have
\begin{align}\label{30}
\frac{{\bf{p}}.\bm{\chi}}{\tau} = \Bigg[\frac{p_0}{T}\partial_0 T + \left(\frac{p^0 - h}{p^0}\right)\frac{\textbf{p}}{T}.\left(\bm{\nabla} T - \frac{T}{nh}\bm{\nabla} P\right)+ T\partial_0\Big(\frac{\mu}{T}\Big)
-\frac{p^jp^{\nu}}{p_0}(\partial_ju_{\nu})- \frac{2q}{p_0}({\bf{E.p}}) - q({\bf{v\times B}}).\bm{\chi}\Bigg].
\end{align}
The derivative is split up covariantly into time and space parts: $\partial_{\mu}= u_{\mu}D + \nabla_{\mu}$, where
		$D = u^{\mu}\partial_{\mu} = (\partial_t,0)$ and $\nabla_{\mu} = \partial_{\mu} - u_{\mu}D = (0,\partial_i)$. The time derivative term ($\partial_0 T$, $\partial_0\left(\frac{\mu}{T}\right))$ can be written in terms of $\nabla_{\mu}u^{\mu}$ by using the following relations \cite{kajantie,s_groot}
		\begin{align}
			\frac{DT}{T} = -\left(\frac{\partial P}{\partial e}\right)_n\nabla_{\mu}u^{\mu},\\
			T D\left(\frac{\mu}{T}\right) = -\left(\frac{\partial P}{\partial n}\right)_e\nabla_{\mu}u^{\mu}.
		\end{align}
		Therefore,
		Eq.\eqref{30} becomes 
		\begin{align}
			\frac{{\bf{p}}.\bm{\chi}}{\tau} = &\Bigg[-p_0\left(\frac{\partial P}{\partial e}\right)_n\nabla_{\mu}u^{\mu} + \left(\frac{p^0 - h}{p^0}\right)\frac{\textbf{p}}{T}.\left(\bm{\nabla} T - \frac{T}{nh}\bm{\nabla} P\right) - \left(\frac{\partial P}{\partial n}\right)_e\nabla_{\mu}u^{\mu}\\\nonumber
			&-\frac{p^jp^{\nu}}{p_0}(\partial_ju_{\nu})- \frac{2q}{p_0}({\bf{E.p}}) - q({\bf{v\times B}}).\bm{\chi}\Bigg].
\end{align}
Since, we are concentrating on thermal transport 
only in presence of magnetic field with no electric current. Therefore, taking the effects of terms associated with
thermal driving forces that corresponds to 
thermal transport in weakly magnetized medium. We have,
\begin{equation}\label{87}
\frac{p^0}{\tau}{\textbf{v}.\left(a_1{\bf{c}} + a_2{\bf{Y}} + a_3({\bf{Y}}\times{\bf{c}})\right)} = \left(p^0 -h\right)\textbf{v}.\textbf{Y}-q\left({\bf{v\times B}}\right).a_2\textbf{Y}-q\left({\bf{v\times B}}\right).a_3\left(\textbf{Y}\times\textbf{c}\right),
\end{equation}
where $\left(\textbf{v}\times\textbf{B}\right).\textbf{c}=0$.  
Using the properties of scalar triple product, 
the parameters $a_1,a_2$ and $a_3$ can be obtained 
by comparing the terms with different 
tensor structures in both sides of Eq.(\ref{87}) independently, 
and we have,
\begin{align}
&\frac{\varepsilon}{\tau}a_1 = a_3 q |{\bf{B}}| ({\bf{c.Y}}),\\
&\frac{\varepsilon}{\tau}a_2 = (\varepsilon-h)-a_3q |{\bf{B}}|,\\
&\frac{\varepsilon}{\tau}a_3 = a_2 q |{\bf{B}}|.
\end{align}
where $p^0\approx\varepsilon, |\textbf{B}|=B$.
 Employing the above equations and defining 
 $\omega_c=\frac{qB}{\varepsilon}$, the parameters 
 reduced to the following forms,
\begin{align}\label{34}
&a_1 = \frac{\tau^3}{\varepsilon}\frac{(\varepsilon-h)}{(1+\omega_c^2\tau^2)}\omega_c^2({\bf{c.Y}}),\nonumber\\
&a_2 = \frac{\tau}{\varepsilon}\frac{(\varepsilon-h)}{(1+\omega_c^2\tau^2)},\nonumber\\
&a_3 = \frac{\tau^2}{\varepsilon}\frac{(\varepsilon-h)}{(1+\omega_c^2\tau^2)}\omega_c.
\end{align}
Substituting $a_1, a_2, a_3$ in Eq.(\ref{29}), 
we obtain $\delta f$ correction to the 
distribution function in the presence of weak 
magnetic field from Eq.(\ref{28}) as,
\begin{equation}\label{37}
\delta f = \frac{\tau(\varepsilon-h)}{(1+\omega_c^2\tau^2)}\left[{\bf{v.Y}}+\tau\omega_c{\bf{v}}.({\bf{Y\times c}}) + \tau^2\omega_c^2({\bf{c.Y}})({\bf{v.c}})\right]\frac{\partial f_0}{\partial\varepsilon}.
\end{equation}
Similarly, $\delta\bar{f}$ can be calculated as
\begin{equation}\label{38}
\delta\bar{f} = \frac{\tau(\varepsilon+\bar{h})}{(1+\omega_c^2\tau^2)}\left[{\bf{v.Y}}-\tau\omega_c{\bf{v}}.({\bf{Y\times c}}) + \tau^2\omega_c^2({\bf{c.Y}})({\bf{v.c}})\right]\frac{\partial\bar{f_0}}{\partial\varepsilon},
\end{equation}
where $\bar{h}$ is the enthalpy per particle for antiquarks. Using Eq.(\ref{37}) and (\ref{38}) in (\ref{20}), 
the heat flow in weakly magnetized medium, 
generalizing to system of different charged 
particles takes the form as
\begin{align}
{\bf{Q}} = &\sum_f g_f\tau_f\int\frac{d^3p}{(2\pi)^3}\frac{{\bf{p}}}{\varepsilon_f}\Bigg[\frac{(\varepsilon_f - h_f)^2}{(1+\omega_c^2\tau_f^2)}\Big({\bf{v.Y}}+\tau_f\omega_c{\bf{v}}.({\bf{Y\times c}}) + \nonumber\\ & \tau_f^2\omega_c^2({\bf{c.Y}})({\bf{v.c}})\Big)\frac{\partial f^0_f}{\partial\varepsilon_f} + \frac{(\varepsilon_f + \bar{h}_f)^2}{(1+\omega_c^2\tau_f^2)}\Big({\bf{v.Y}}-\tau_f\omega_c{\bf{v}}.({\bf{Y\times c}}) + \nonumber\\ & \tau_f^2\omega_c^2({\bf{c.Y}})({\bf{v.c}})\Big)\frac{\partial \bar{f}^0_f}{\partial\epsilon_f}\Bigg],
\end{align}
where $h_f$ and $\bar{h}_f$ is the enthalpy per particle of quarks and antiquarks for $f$-th flavor respectively. Simplifying the analysis by fixing the direction 
of ${\bf{B}}$ along z-axis and temperature gradient 
in x-y plane. Under this condition, heat flow 
assumes the form
\begin{equation}
{\bf{Q}} = -\kappa_0 T{\bf{Y}} - \kappa_1 T({\bf{Y\times c}}),
\end{equation}
where thermal transport coefficients in weakly 
magnetized medium, $\kappa_0 (=\kappa_q + \kappa_g)$ and $\kappa_1$, can 
be defined as,
\begin{equation}\label{41}
\kappa_0 = \sum_f\frac{g_f\tau_f}{6\pi^2 T^2}\int dp\frac{p^4}{\varepsilon_f^2}\Bigg[\frac{(\varepsilon_f - h_f)^2}{(1+\omega_c^2\tau_f^2)}f_f^0(1-f_f^0) + \frac{(\varepsilon_f + \bar{h}_f)^2}{(1+\omega_c^2\tau_f^2)}\bar{f}_f^0(1-\bar{f}_f^0)\Bigg] + \kappa_g,
\end{equation}
and
\begin{equation}\label{42}
\kappa_1 = \sum_f\frac{g_f\tau_f^2}{6\pi^2 T^2}\int dp\frac{p^4}{\varepsilon_f^2}\Bigg[\frac{(\varepsilon_f - h_f)^2\omega_c}{(1+\omega_c^2\tau_f^2)}f_f^0(1-f_f^0) - \frac{(\varepsilon_f + \bar{h}_f)^2\omega_c}{(1+\omega_c^2\tau_f^2)}\bar{f}_f^0(1-\bar{f}_f^0)\Bigg],
\end{equation}
where $f$ stands for $f$th flavor. Similar to the discussion of charge transport 
coefficients, here we have thermal ($\kappa_{0}$) 
and Hall-type thermal conductivity ($\kappa_{1}$). Hall-type thermal 
conductivity emerges due to the transverse temperature 
gradient which is induced by the action of 
magnetic field perpendicular to initial direction 
of heat current and would be contributed by quarks and anti-quarks. So far, we have obtained the contribution due to quarks and anti-quarks to the thermal conductivity. Now, we will calculate the gluonic contribution to thermal conductivity. Since gluons do not interact with electromagnetic field therefore Eq.\eqref{23} for gluons assumes the form as
\begin{equation}\label{gluon_eq}
	p^{\mu}\partial_{\mu}T\left(\frac{\partial f_g}{\partial T}\right) + 
	 p^{\mu}\partial_{\mu} (p^{\nu}u_{\nu})\left(\frac{\partial f_g}{\partial p^0_g}\right)
= -\frac{p^{\mu}u_{\mu}}{\tau}\delta f_g,
\end{equation}
where $ f_g = 1/\left(e^{\varepsilon_g/T}-1\right),p^0_g = \varepsilon_g$ is the single particle energy of gluon. Using the following partial derivative
\begin{equation}
	\frac{\partial f_g}{\partial T} = \frac{\varepsilon}{T^2}f_g(1+f_g),
\end{equation}
\begin{equation}
	\frac{\partial f_g}{\partial p^0_g} = -\frac{1}{T}f_g(1+f_g),
\end{equation}
the Eq.\eqref{gluon_eq} can be rewritten as
\begin{equation}
	\frac{1}{T}f_g\left(1+f_g\right)\left[\frac{1}{T}\left(p^0_g DT+p^j\nabla_jT\right)+T D\left(\frac{\mu}{T}\right)+\frac{p^jT}{p^0_g}\nabla_j\left(\frac{\mu}{T}\right)-\frac{1}{p^0_g}\left(p^0_gp^{\nu}\partial_{0}u_{\nu}+p^jp^{\nu}\partial_j u_{\nu}\right)\right] = -\frac{\delta f_g}{\tau_g}.
\end{equation}
Considering only thermal driving forces and taking the gluon chemical potential to be zero, we obtain the equation as
\begin{equation}
	\frac{1}{T}f_g\left(1+f_g\right)\left[p^j\left(\frac{\nabla_jT}{T}-\frac{\nabla_j P}{nh}\right)\right] = -\frac{\delta f_g}{\tau_g},
\end{equation}
where $\tau_g$ is thermal relaxation time for gluons is given as \cite{kajantie}
\begin{equation}
	\tau_g = \frac{1}{22.5 T\alpha_s^2\log\left(\frac{1}{\alpha_s}\right)(1+0.06 N_f)}.
\end{equation}
Putting $\delta f_g$ in Eq.\eqref{20} and comparing with Eq.\eqref{71}, we obtain the gluonic contribution to the thermal conductivity as
\begin{equation}
	\kappa_g = \frac{g_g\tau_g}{6\pi^2T^2}\int dp \quad p^4 f_g\left(1+f_g\right).
\end{equation}
Therefore, thermal conductivity due to quarks, antiquarks and gluons along the initial direction of heat current is given by
\begin{equation}\label{total_therm}
\kappa_{0}=	\sum_f\frac{g_f\tau_f}{6\pi^2 T^2}\int dp\frac{p^4}{\varepsilon_f^2}\Bigg(\frac{(\varepsilon_f - h_f)^2}{(1+\omega_c^2\tau_f^2)}f_f^0(1-f_f^0) + \frac{(\varepsilon_f + \bar{h}_f)^2}{(1+\omega_c^2\tau_f^2)}\bar{f}_f^0(1-\bar{f}_f^0)\Bigg) + \frac{g_g\tau_g}{6\pi^2T^2}\int dp\quad p^4 f_g\left(1+f_g\right).
\end{equation}
Gluons will not contribute to $\kappa_{1}$ because the Lorentz force will not change their direction of motion.
 The thermal 
conductivity is obtained from heat current in 
temperature gradient on the condition that 
there is no electric current \cite{haug}. 
 Hall-type thermal conductivity is the thermal analog of 
classical Hall effect where temperature plays 
the role of voltage and heat flow replaces the 
electric current \cite{charles} and it is the 
Lorentz force acting on charged particles affecting 
the curvature of carrier's trajectories through
 the magnetic field. At zero chemical potential, $\kappa_{1}$ will not vanish due to the unequal contribution from quarks and anti-quarks in the same direction, unlike in the case of Hall conductivity in charge transport. 
\section{RESULTS AND DISCUSSIONS} \label{results}
In this section, we will discuss the results regarding 
the Ohmic and Hall conductivity, thermal and 
Hall-type thermal conductivity and further 
Knudsen number and Wiedemann-Franz law as their application.
\subsection{Ohmic and Hall conductivity}
\begin{figure}[H]
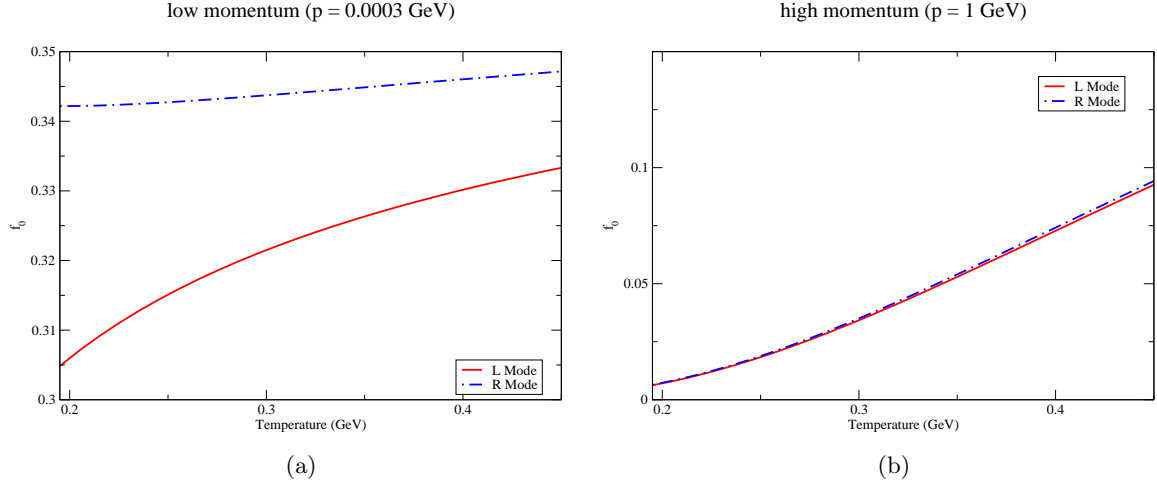

	\begin{subfigure}{0.48\textwidth}
		\includegraphics[width=0.95\textwidth]{f_c_p1.eps}
		\caption{}
	\end{subfigure}
	\begin{subfigure}{0.48\textwidth}
		\includegraphics[width=0.95\textwidth]{f_c_p2.eps}
		\caption{}
	\end{subfigure}
	\caption{Variation of distribution function of quark ($f_0$) in left- and right-handed mode with temperature, where effective quark mass has been used.}\label{dist_p}
\end{figure}
\begin{figure}[H]
	\begin{subfigure}{0.48\textwidth}
		\includegraphics[width=0.95\textwidth]{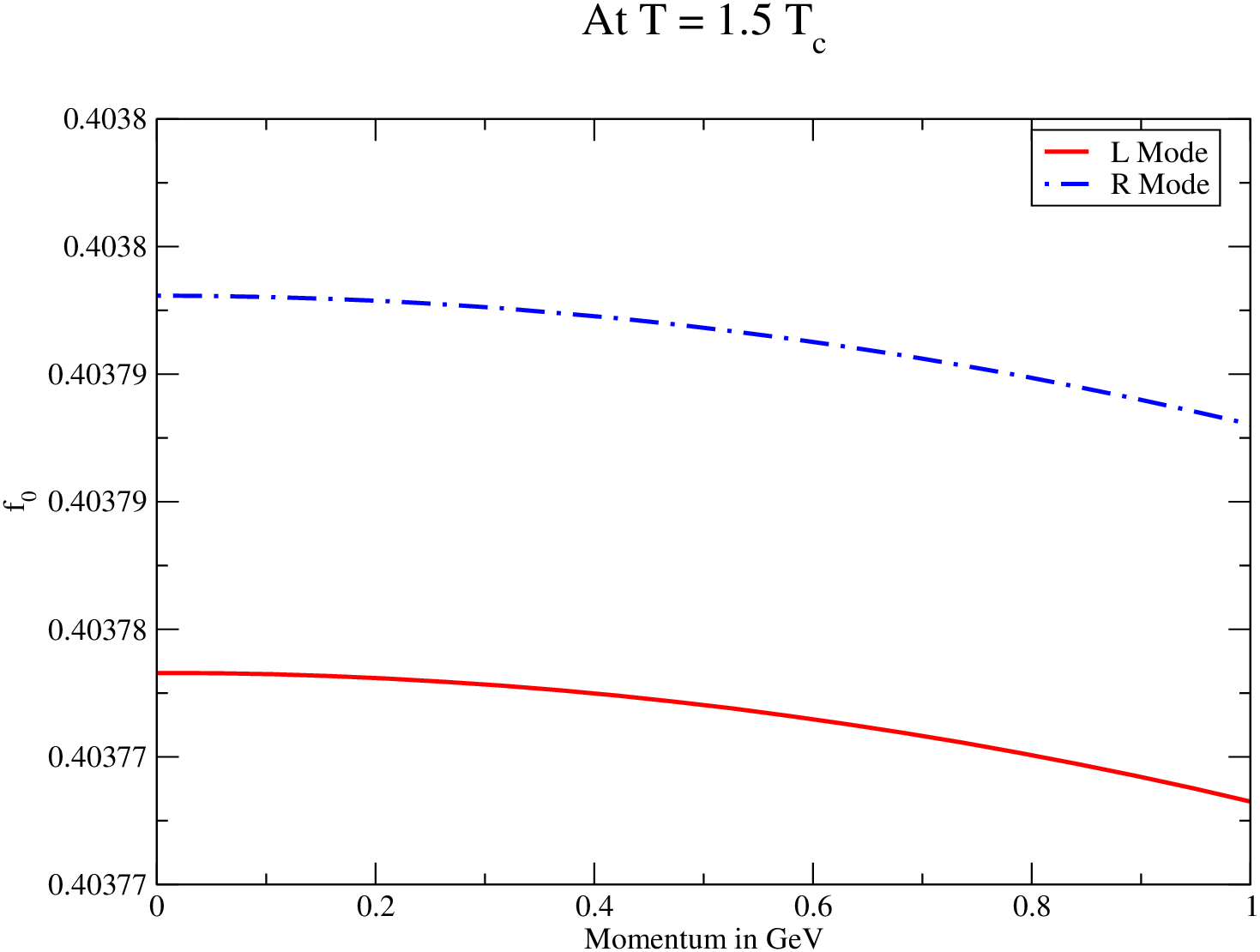}
		\caption{}
	\end{subfigure}
	\begin{subfigure}{0.48\textwidth}
		\includegraphics[width=0.95\textwidth]{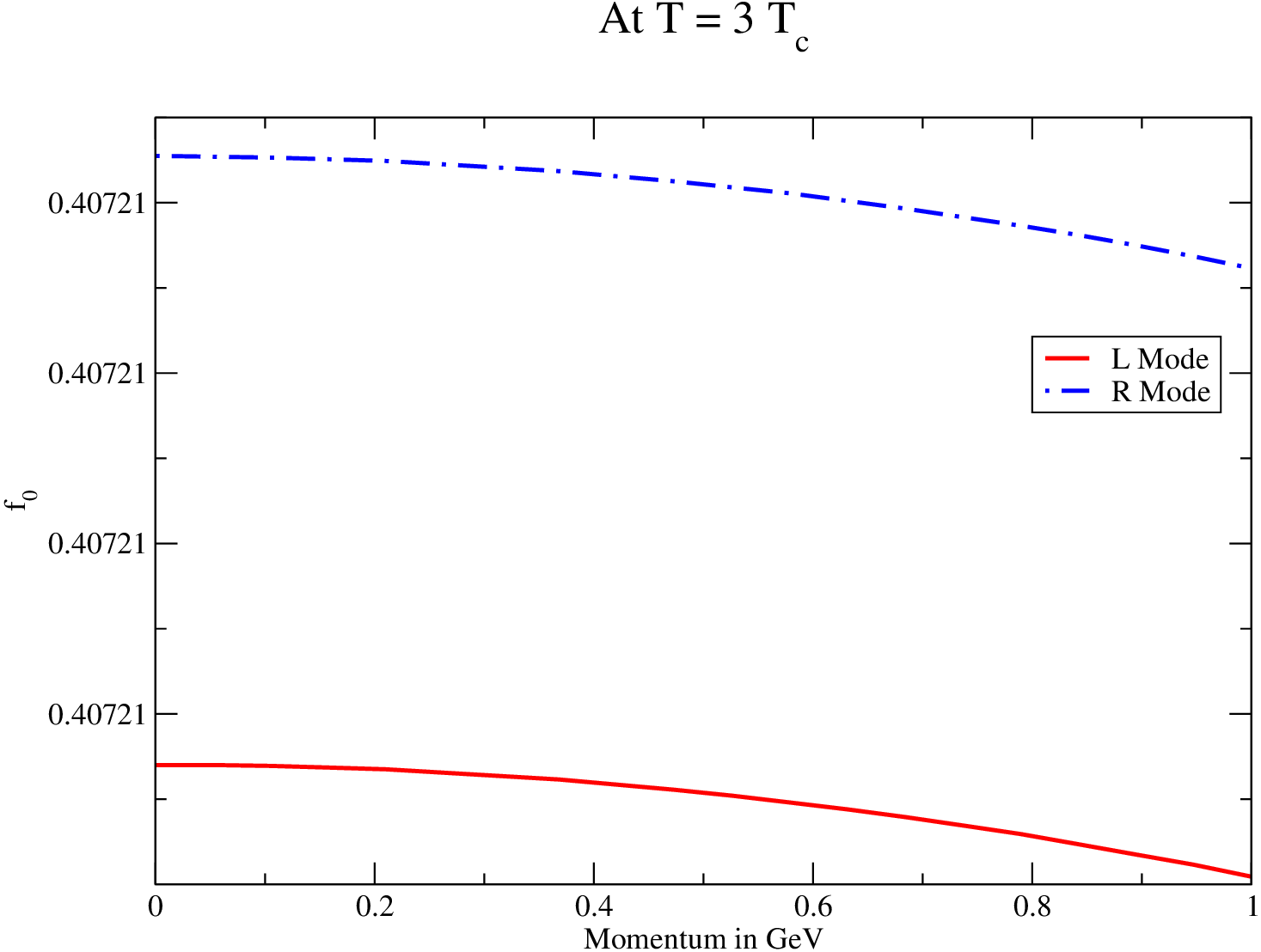}
		\caption{}
	\end{subfigure}
	\caption{Variation of distribution function of quark ($f_0$) in left- and right-handed mode with momentum, where effective quark mass has been used.}\label{dist_t}
\end{figure}
\begin{figure}[H]
	\begin{subfigure}{0.48\textwidth}
		\includegraphics[width=0.95\textwidth]{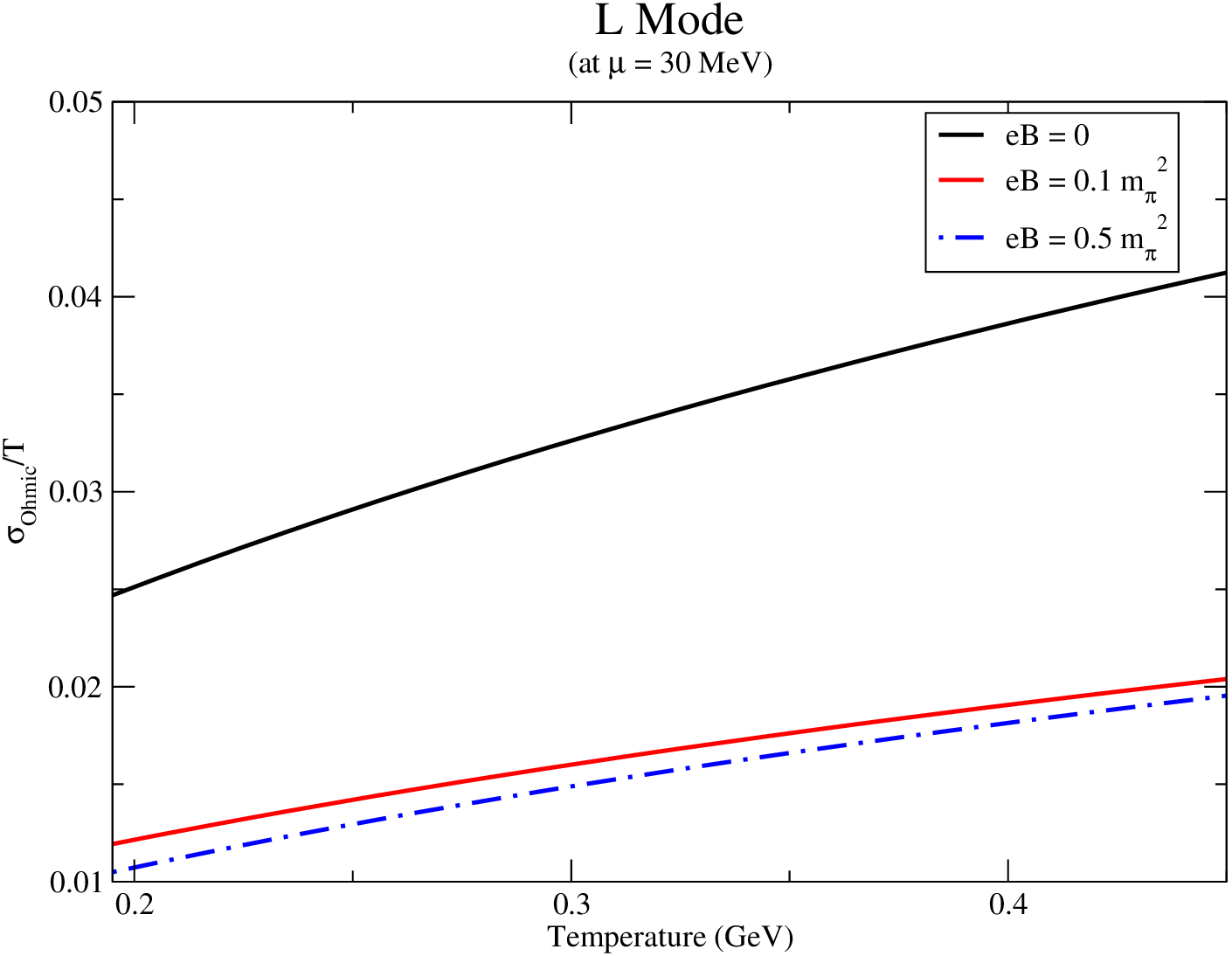}
		\caption{}\label{fig1a}
	\end{subfigure}
	\begin{subfigure}{0.48\textwidth}
		\includegraphics[width=0.95\textwidth]{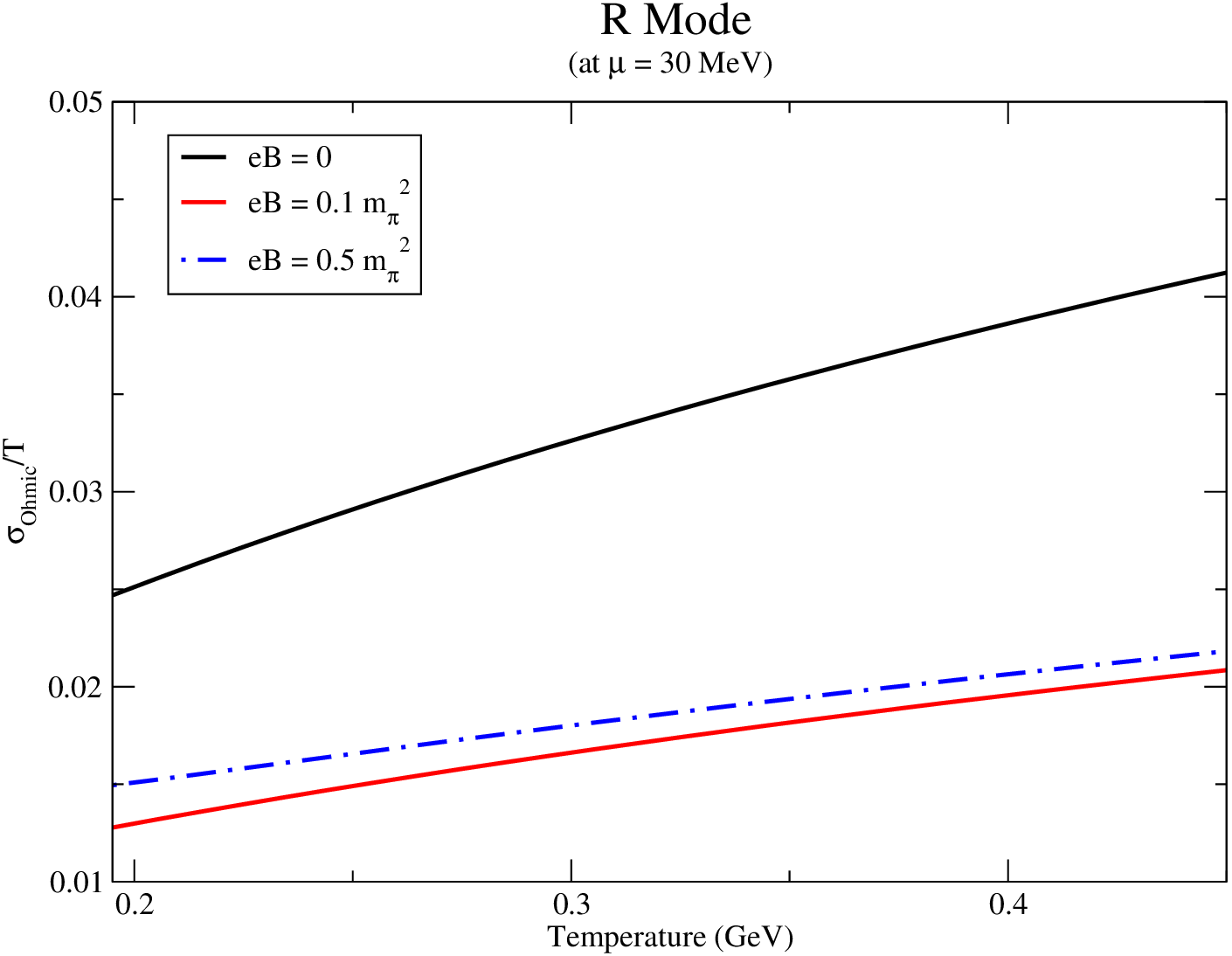}
		\caption{}\label{fig1b}
	\end{subfigure}
	\caption{Variation of $\sigma_{\text{Ohmic}}/T$ for L mode and R mode with temperature at different fixed values of magnetic field.}\label{fig1}
\end{figure}
\begin{figure}[H]
	\begin{subfigure}{0.48\textwidth}
		\includegraphics[width=0.95\textwidth]{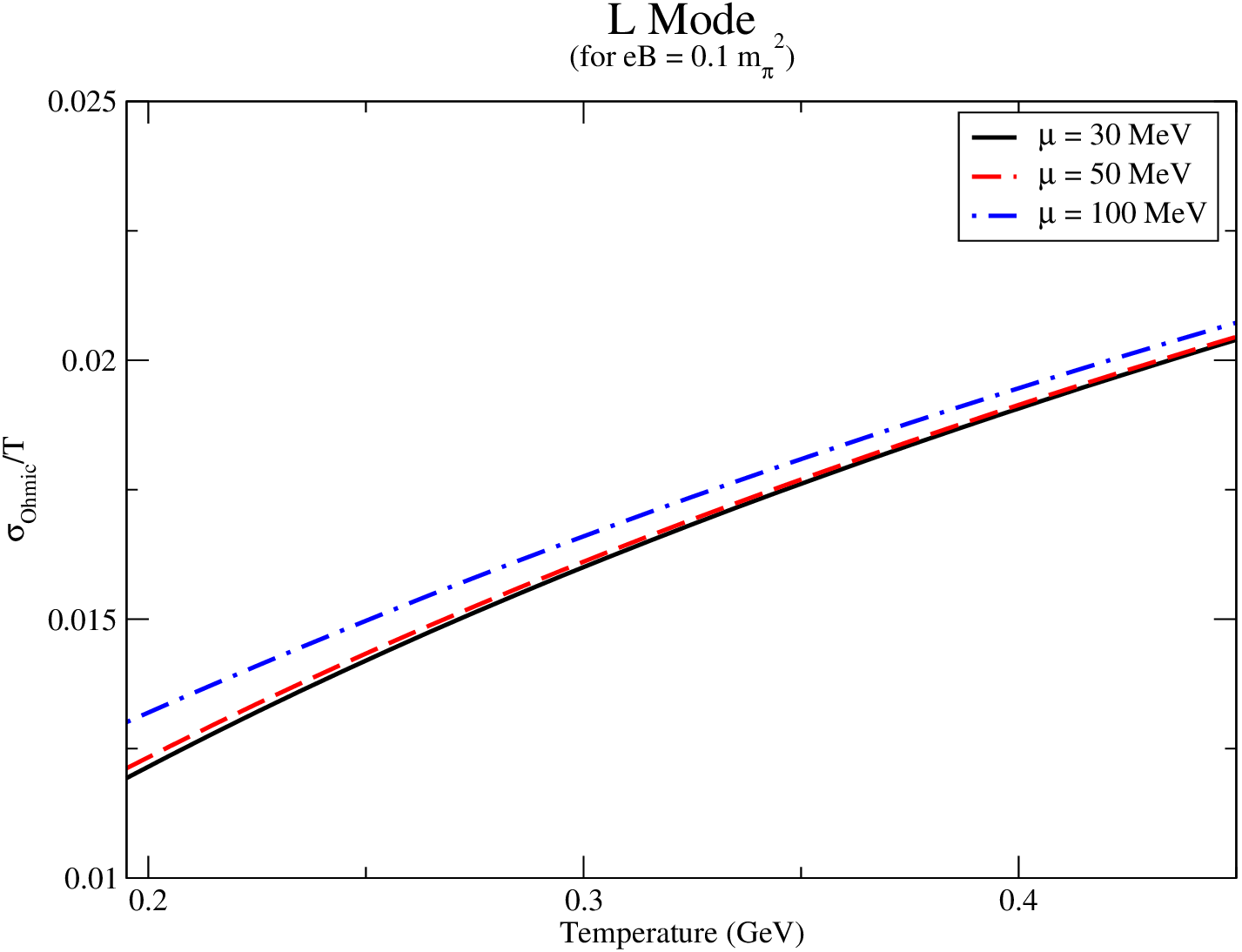}
		\caption{}\label{fig2a}
	\end{subfigure}
	\begin{subfigure}{0.48\textwidth}
		\includegraphics[width=0.95\textwidth]{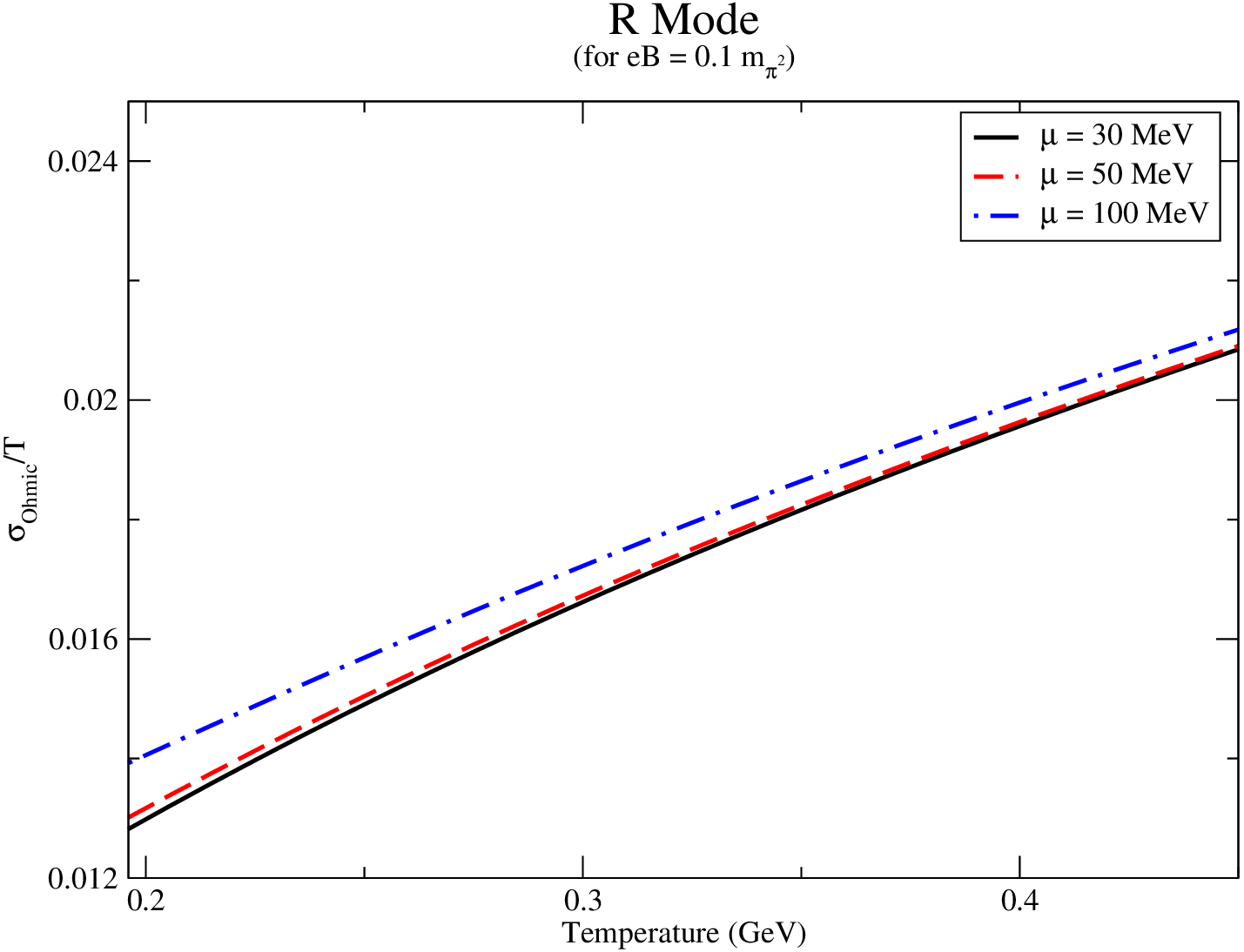}
		\caption{}\label{fig2b}
	\end{subfigure}
	\caption{Variation of $\sigma_{\text{Ohmic}}/T$ for L mode and R mode with temperature at different fixed values of quark chemical potential.}\label{fig2}
\end{figure}

\begin{figure}[H]
	\begin{subfigure}{0.48\textwidth}
		\includegraphics[width=0.95\textwidth]{hall_l.eps}
		\caption{}
	\end{subfigure}
	\begin{subfigure}{0.48\textwidth}
		\includegraphics[width=0.95\textwidth]{hall_r.eps}
		\caption{}
	\end{subfigure}
	\caption{Variation of $\sigma_{\text{Hall}}/T$ for L mode and R mode with temperature at different fixed values of magnetic field.}\label{hall}
\end{figure}
\begin{figure}[H]
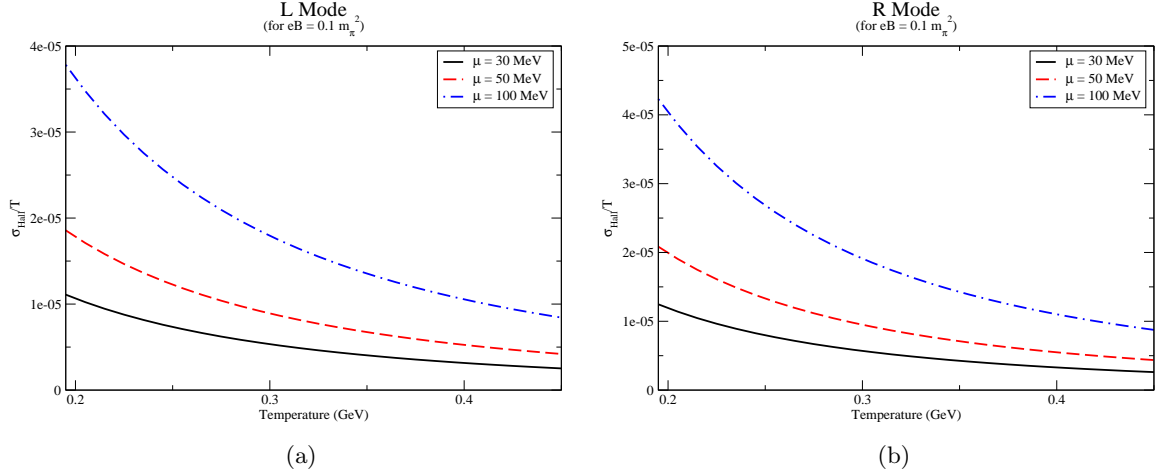

	\begin{subfigure}{0.48\textwidth}
		\includegraphics[width=0.95\textwidth]{hall_mu_l.eps}
		\caption{}
	\end{subfigure}
	\begin{subfigure}{0.48\textwidth}
		\includegraphics[width=0.95\textwidth]{hall_mu_r.eps}
		\caption{}
	\end{subfigure}
	\caption{Variation of $\sigma_{\text{Hall}}/T$ for L mode and R mode with temperature at different fixed values of quark chemical potential.}\label{hall_mu}
\end{figure}
In Fig.\eqref{fig1}, we have shown the variation of 
ratio of Ohmic conductivity to temperature 
($\sigma_{{\text{Ohmic}}}/T$) with respect to 
temperature for zero and finite
 magnetic field at non-zero chemical potential 
 ($\mu$=30 MeV). It is evident that the magnitude of $\sigma_{{\text{Ohmic}}}/T$ gets decrease 
 in presence of magnetic field as shown in 
 Fig.\eqref{fig1a} and \eqref{fig1b}. The difference between the magnitude of conductivities for L- and R-mode increases with increase in magnetic field due different effective quark mass for L- and R-mode Eq.\eqref{qaurk_mass}. This can also be deduced from the plots of distribution function of quark for 
 left-handed and right-handed mode at 
 fixed momentum and temperature shown in Fig.\eqref{dist_p} and \eqref{dist_t} respectively where $T_c \simeq 160 MeV$ is the deconfinement temperature from hadron phase to QGP phase. Further, 
 $\sigma_{{\text{Ohmic}}}/T$ for 
  L mode decreases with magnetic field whereas it 
  shows increasing trend for R mode. This behaviour of $\sigma_{{\text{Ohmic}}}/T$ with magnetic 
  field for L and R mode is 
  attributed to the factor $\frac{\tau}{\varepsilon^2(1+\omega_c^2\tau^2)}$.
   The thermal mass squared with magnetic field 
  correction for left (right) handed mode 
  is $m_{L(R)}^2 = m_{th}^2 \pm 4g^2 C_F M^2$, which
   is found to increase (decreasing) with magnetic 
   field. As this mass appears in the denominator of $\frac{\tau}{\varepsilon^2(1+\omega_c^2\tau^2)}$ 
   which leads to the decreasing (increasing) 
   behaviour of $\sigma_{{\text{Ohmic}}}/T$ for 
 left (right) handed mode. The increasing behaviour of $\sigma_{\text{Ohmic}}/T$ with temperature for 
  both modes could be due to the Boltzmann 
   factor $\exp(-\varepsilon(p)/T)$ in the 
distribution function. Fig.\eqref{fig2} shows 
the variation of normalized Ohmic conductivity 
at different constant values of quark chemical potential for
 eB = $ 0.1m_{\pi}^2$,
where it increases with increase in quark chemical 
potential for both modes. With increasing quark 
chemical potential the Boltzmann factor $\exp(-\mu/T)$ 
increases due to the higher contribution 
from quarks than antiquarks. 
The transverse motion of charged 
particle under the action of 
Lorentz force leads to the generation of
 Hall current. The variation of 
 $\sigma_{{\text{Hall}}}/T$ with temperature for 
different values of magnetic field at $\mu = 30$ 
MeV is shown in  Fig.\eqref{hall} for both modes. 
It increases with magnetic field as $\sigma_{{\text{Hall}}}/T$ 
is proportional to $\omega_c$ for the considered 
range of temperature, chemical potential and 
magnetic field. The decreasing behaviour of 
normalized Hall conductivity with temperature 
is predominantly due to the factor 
$\omega_c\tau$ in the numerator of 
Eq.\eqref{16}. $\sigma_{{\text{Hall}}}/T$ 
for left-handed mode is relatively smaller 
than right-handed mode as the mass for left 
mode is comparatively larger than right mode. 
Hence, we can say that the variation of 
Ohmic conductivity with magnetic field 
is affected through the effective mass 
as shown in Fig.\eqref{fig1}, where at $eB = 0$, $\sigma_{{\text{Ohmic}}}/T$ has relatively
 higher magnitude. At $eB = 0$, Hall 
 conductivity vanishes and its behaviour 
 with magnetic field is affected through 
 the direct dependence on $q_f B$ in the 
 numerator of Eq.\eqref{16}. Similar to Ohmic 
conductivity, Hall conductivity also increases 
with quark chemical potential for L- and R-mode 
as shown in Fig.\eqref{hall_mu}. At zero chemical
 potential, number of quarks and antiquarks are 
 same and their contribution to the Hall current 
 is same but opposite in direction. So, the net 
 Hall current vanishes at zero chemical potential
  and can be explicitly seen in Eq. \eqref{16}.\\
\subsection{Thermal and Hall-type thermal conductivity}
\begin{figure}[H]
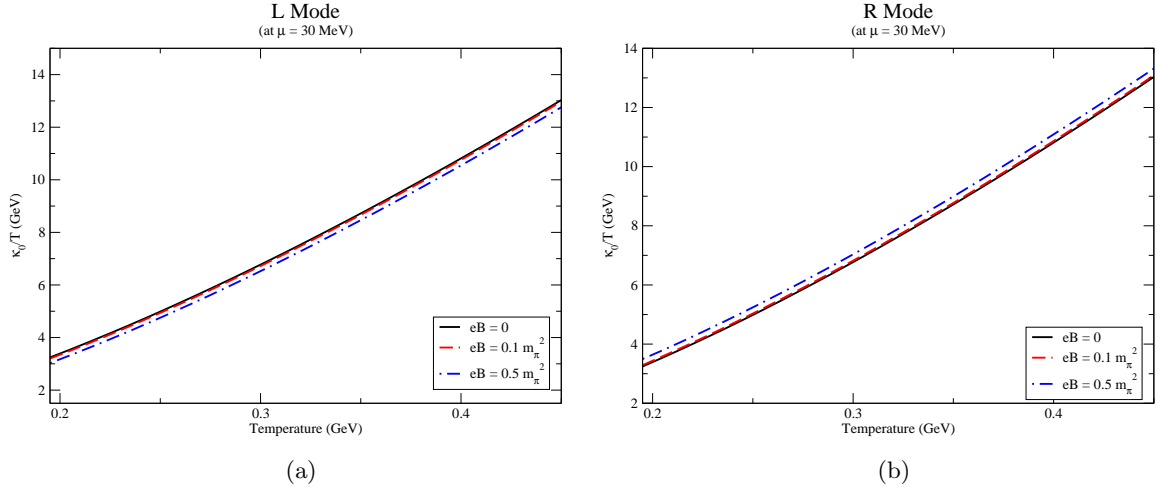

	\begin{subfigure}{0.48\textwidth}
		\includegraphics[width=0.95\textwidth]{kappa_0_l.eps}
		\caption{}\label{fig3a}
	\end{subfigure}
	\begin{subfigure}{0.48\textwidth}
		\includegraphics[width=0.95\textwidth]{kappa_0_r.eps}
		\caption{}\label{fig3b}
	\end{subfigure}
	\caption{Variation of $\kappa_{0}/T$ for L and R mode with temperature at different fixed values of magnetic field.}\label{fig3}
\end{figure}
\begin{figure}[H]
	\begin{subfigure}{0.48\textwidth}
		\includegraphics[width=0.95\textwidth]{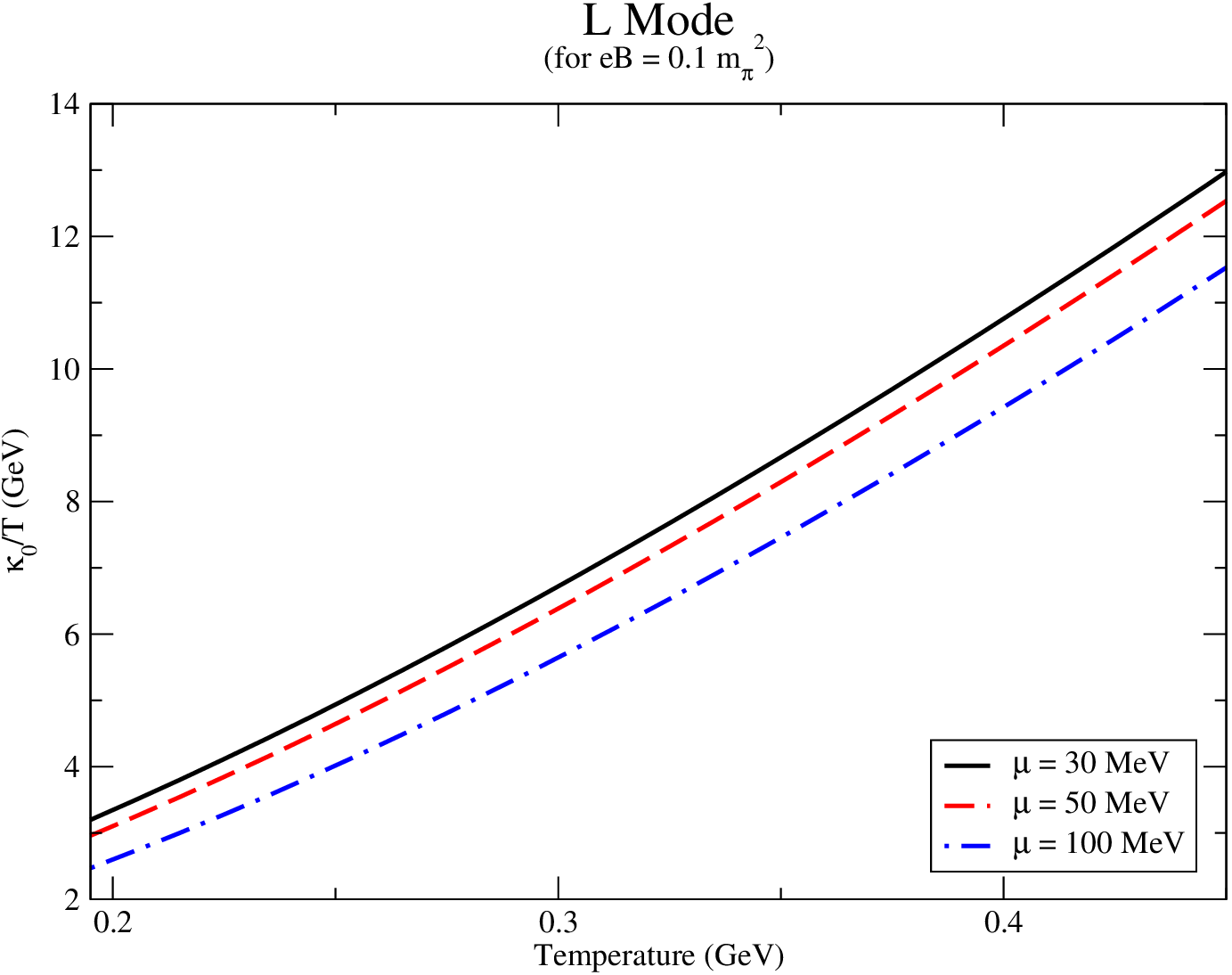}
		\caption{}
	\end{subfigure}
	\begin{subfigure}{0.48\textwidth}
		\includegraphics[width=0.95\textwidth]{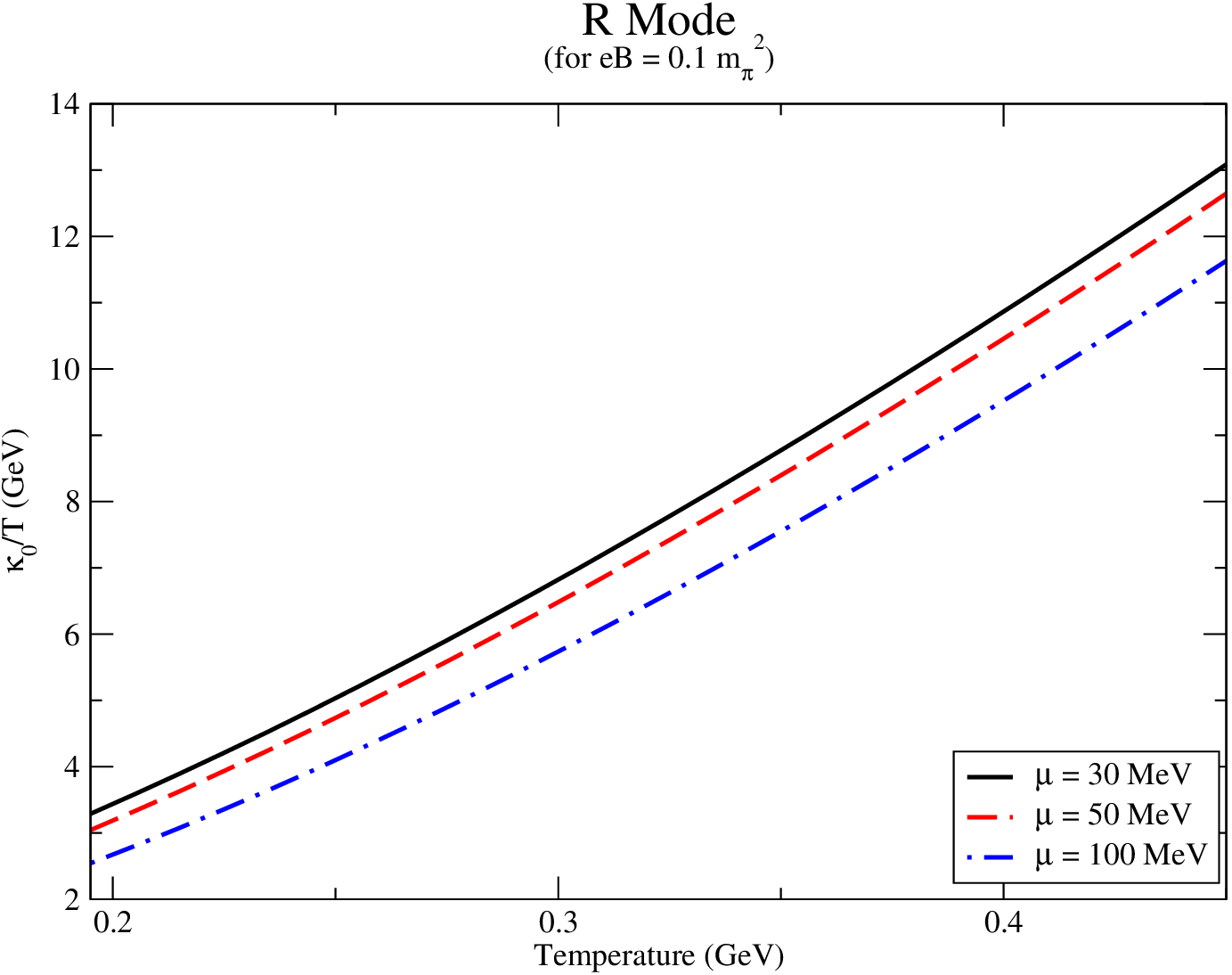}
		\caption{}
	\end{subfigure}
	\caption{Variation of $\kappa_{0}/T$ for L and R mode with temperature at different fixed values of quark chemical potential.}\label{k_0_poten}
\end{figure}
\begin{figure}[H]
	\begin{subfigure}{0.48\textwidth}
		\includegraphics[width=0.95\textwidth]{kappa_1_l.eps}
		\caption{}\label{fig4a}
	\end{subfigure}
	\begin{subfigure}{0.48\textwidth}
		\includegraphics[width=0.95\textwidth]{kappa_1_r.eps}
		\caption{}\label{fig4b}
	\end{subfigure}
	\caption{Variation of $\kappa_{1}/T$ for L and R mode with temperature at different fixed values of magnetic field.}\label{fig4}
\end{figure}
\begin{figure}[H]
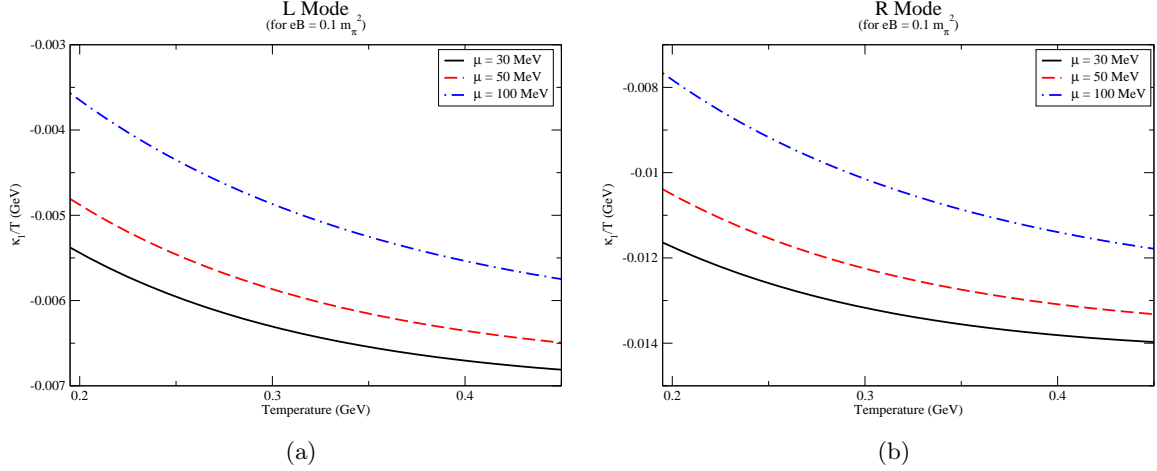

	\begin{subfigure}{0.48\textwidth}
		\includegraphics[width=0.95\textwidth]{kappa_1_mu_l.eps}
		\caption{}
	\end{subfigure}
	\begin{subfigure}{0.48\textwidth}
		\includegraphics[width=0.95\textwidth]{kappa_1_mu_r.eps}
		\caption{}
	\end{subfigure}
	\caption{Variation of $\kappa_{1}/T$ for L and R mode with temperature at different fixed values of quark chemical potential.}\label{k_1_mu}
\end{figure}
Fig.\eqref{fig3} and \eqref{k_0_poten} shows the 
variation of ratio of thermal conductivity to 
temperature ($\kappa_{0}/T$) for left and 
right-handed mode with temperature at different 
fixed values of magnetic field and quark 
chemical potential respectively. At 
zero magnetic field, there would be 
no lifting of degeneracy and hence 
we compared $\kappa_{0}/T$ in the 
absence and presence of magnetic 
field (with both modes). $\kappa_{0}/T$ 
increases with temperature for both modes and 
has approximately the same value at eB = 0.1 $m_{\pi}^2$. 
The increasing behaviour of $\kappa_{0}/T$ with temperature is 
due to the factor $(\varepsilon-h)^2$, \rm{$(\varepsilon+h)^2$} and 
distribution function as can be seen from Eq.\eqref{total_therm}. Further, $\kappa_{0}/T$ for 
L mode decreases with magnetic field whereas for
 R mode it increases with magnetic 
 field. The difference between magnitude
  of thermal conductivity for left- and 
right-handed mode is again attributed 
to the different effective quark mass 
for both modes, similar to the $\sigma_{{\text{Ohmic}}}/T$.
 Since, $(\varepsilon+h)^2$ is higher in magnitude than $(\varepsilon-h)^2$ therefore $(\varepsilon+h)^2\exp(\mu/T)$ leads to the decreasing behaviour of $\kappa_{0}/T$ with quark chemical potential for both modes.
Furthermore, due to the 
 deflected motion of particles under the action 
 of Lorentz force, there is generation of Hall 
 component of thermal conductivity ($\kappa_{1}$) 
 in a direction perpendicular to both the magnetic field 
 and initial thermal driving force. The variation 
 of $\kappa_{1}/T$ with temperature at different 
fixed values of magnetic field and quark chemical 
potential is shown in Fig.\eqref{fig4} 
and \eqref{k_1_mu} respectively for both 
modes. Considering the absolute value of the ratio $(\kappa_{1}/T)$, we infer that $(\kappa_{1}/T)$ increases with temperature and magnetic field. The increasing behaviour with temperature is due to the $(\varepsilon+h)^2$ factor in the numerator of Eq.\eqref{42}. Moreover, the direct dependence on magnetic field leads to the amplification of Hall-type thermal conductivity with magnetic field. Further, $\kappa_{1}/T$ decreases
with quark chemical potential and will not vanish for $\mu=0$ due to the unequal contribution from quarks and anti-quarks in the same direction. We can also infer that 
the behaviour of longitudinal thermal 
conductivity with magnetic field is 
affected through the effective quark 
mass for both modes whereas Hall type 
thermal conductivity is affected 
through direct dependence on magnetic 
field as could be seen in Eq.\eqref{42}. $\kappa_{1}/T$ is
comparatively smaller in magnitude than $\kappa_{0}/T$,
similar to the charge
 transport.  

\subsection{Knudsen Number}
\begin{figure}[H]
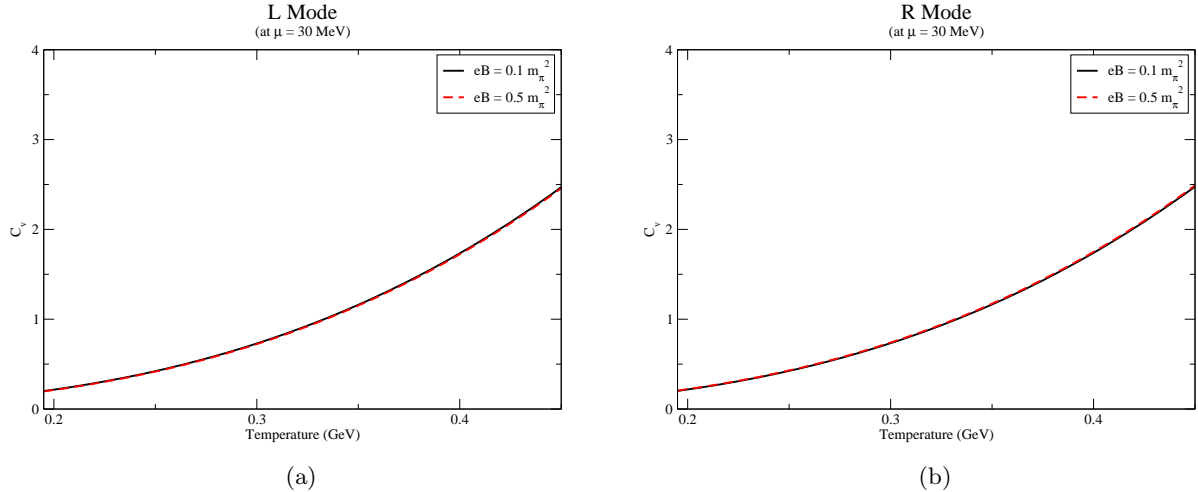

	\begin{subfigure}{0.48\textwidth}
		\includegraphics[width=0.95\textwidth]{C_v_l.eps}
		\caption{}\label{cvl}
	\end{subfigure}
\hspace{5mm}
	\begin{subfigure}{0.48\textwidth}
		\includegraphics[width=0.95\textwidth]{C_v_r.eps}
		\caption{}\label{cvr}
	\end{subfigure}
	\caption{Variation $C_v$ for L and R mode with temperature where effective quark mass has been used for both modes.}\label{cv}
\end{figure}
\begin{figure}[H]
	\begin{subfigure}{0.48\textwidth}
		\includegraphics[width=0.95\textwidth]{knudsen.eps}
		\caption{}\label{fig5a}
	\end{subfigure}
	\begin{subfigure}{0.48\textwidth}
		\includegraphics[width=0.95\textwidth]{knudsen_0_r.eps}
		\caption{}\label{fig5b}
	\end{subfigure}
	\caption{Variation of Knudsen number ($\Omega_0$) for L- and R-mode with temperature at different fixed values of magnetic field.  }\label{fig5}
\end{figure}
\begin{figure}[H]
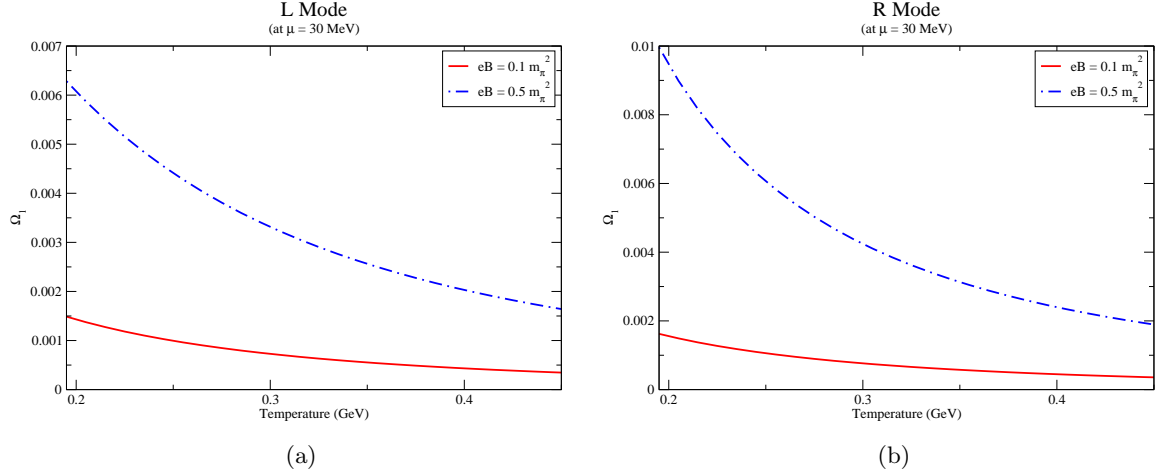

	\begin{subfigure}{0.48\textwidth}
		\includegraphics[width=0.95\textwidth]{knud_1_l.eps}
		\caption{}\label{fig7a}
	\end{subfigure}
	\begin{subfigure}{0.48\textwidth}
		\includegraphics[width=0.95\textwidth]{knudsen_1_r.eps}
		\caption{}\label{fig7b}
	\end{subfigure}
	\caption{Variation of Knudsen number ($\Omega_1$) for L- and R-mode with temperature at different fixed values of magnetic field.  }\label{fig7}
\end{figure}
The applicability of ideal hydrodynamic requires 
local thermal equilibration. The degree of 
thermalization in fluid produced in heavy 
ion collision can be characterized by 
dimensionless parameter which is termed 
as Knudsen number ($\Omega$), which is 
the ratio of microscopic length scale 
(mean free path) to the macroscopic
 length scale (characteristic length scale)
  of the system \cite{chaudhari}. 
  The mean free path ($\lambda$) is 
  identified as $3\kappa/(\mathrm{v} C_v)$ 
  with $\mathrm{v}$ and $C_v$ as relative 
  velocity and specific heat at constant 
  volume respectively. Knudsen number can be recast as
  \begin{equation}
  	\Omega = \frac{3\kappa}{\text{v}C_v L},
  \end{equation}
 where we have taken $\text{v}\simeq1$, 
 $L$=4 \text{fm}. $C_v$ is evaluated from 
 the temperature gradient of energy density, i.e., $C_v = \partial(u_{\alpha}T^{\alpha\beta}u_{\beta})/\partial T$.
  The small value of
 Knudsen number implies the large number 
 of collisions which bring the system back
  to local equilibrium. The behaviour of $\Omega_0$ 
  (associated to $\kappa_{0}$ and $\kappa_g$)
 and $\Omega_1$ (associated to $\kappa_{1}$)
  with magnetic field is found to be
   closely related to behaviour of 
   $\kappa_{0}$ and $\kappa_{1}$ for 
   both modes. As we can see that 
   effect of magnetic field on $C_v$ 
   is not so much pronounced as 
   shown in Fig.\eqref{cvl} and \eqref{cvr}.  $\Omega_0$ 
   decreases with magnetic field 
   for L-mode whereas increases with 
   magnetic field for R-mode as 
   shown in Fig.\eqref{fig5}, similar 
   to the $\kappa_{0}/T$. Moreover, 
   $\Omega_1$ shows the increasing 
   trend with magnetic field as 
   shown in Fig.\eqref{fig7}, similar 
   to the trends followed by $\kappa_{1}/T$ (taking the absolute value of $\kappa_{1}$).
    Knudsen number ($\Omega_0$ and 
    $\Omega_1$) is found to be less 
    than unity for both modes in presence 
 of weak magnetic field
 at $\mu= 30 MeV$, thus ensures the system 
 to be in thermal equilibrium. 
\subsection{Wiedemann-Franz law}
\begin{figure}[H]
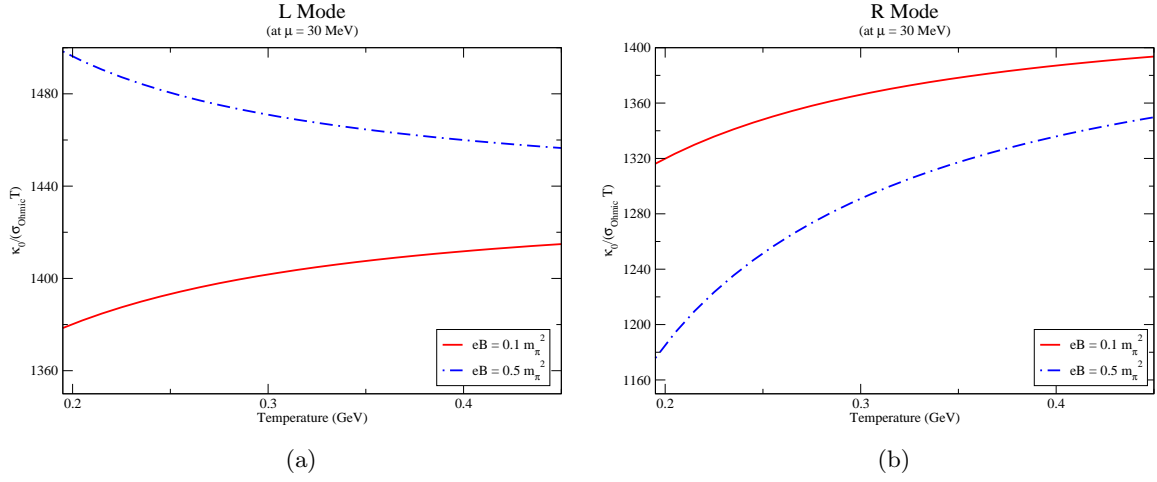

	\begin{subfigure}{0.48\textwidth}
		\includegraphics[width=0.95\textwidth]{lorenz.eps}
		\caption{}\label{fig6a}
	\end{subfigure}
	\begin{subfigure}{0.48\textwidth}
		\includegraphics[width=0.95\textwidth]{lorenz_r.eps}
		\caption{}\label{fig6b}
	\end{subfigure}
	\caption{Variation of Lorenz number for L- and R-mode with temperature at different fixed values of magnetic field. }\label{fig6}
\end{figure}

\begin{figure}[h]
	\centering
	\includegraphics[width=0.45\textwidth]{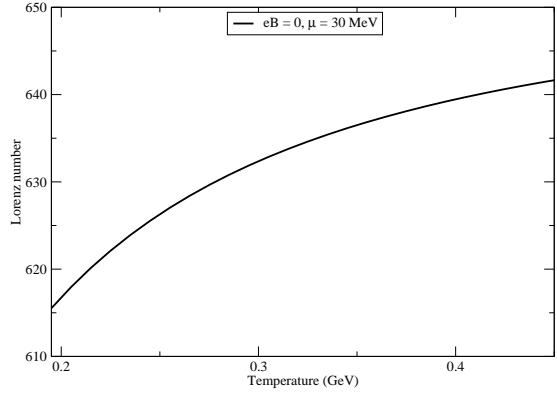}
	\caption{Variation of Lorenz number with temperature for zero magnetic field at finite chemical potential.}\label{lorenz_t}
\end{figure}

\begin{figure}[H]
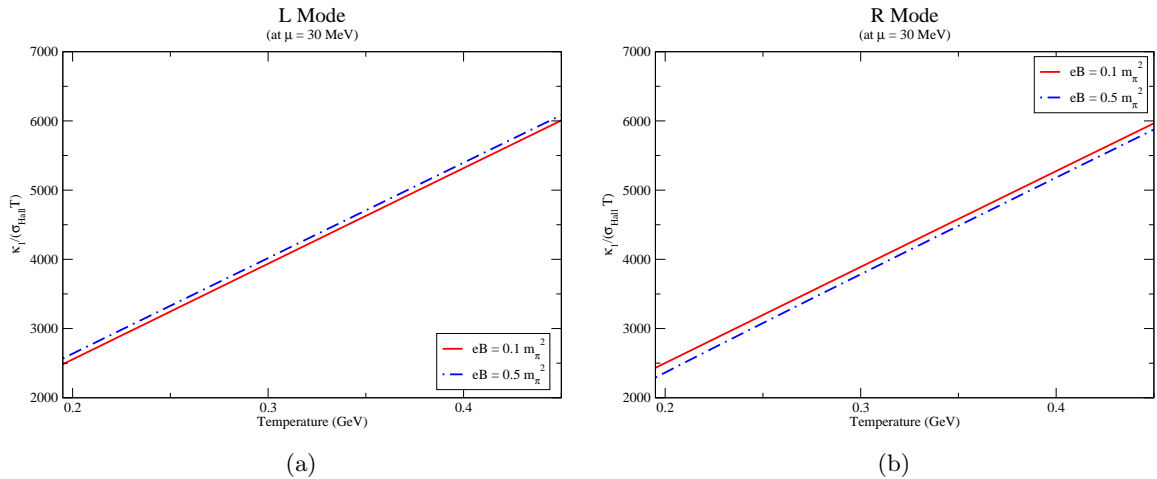

	\begin{subfigure}{0.48\textwidth}
		\includegraphics[width=0.95\textwidth]{lorenz_1.eps}
		\caption{}\label{fig8a}
	\end{subfigure}
	\begin{subfigure}{0.48\textwidth}
		\includegraphics[width=0.95\textwidth]{lorenz_1_r.eps}
		\caption{}\label{fig8b}
	\end{subfigure}
	\caption{Variation of Hall Lorenz number for L- and R-mode with temperature at different fixed values of magnetic field. }\label{fig8}
\end{figure}

The interplay between charge and heat 
transport coefficients can be understood 
via Wiedemann-Franz law. The temperature 
behaviour of Lorenz number ($\kappa_{0}/\sigma_{\text{Ohmic}}T$) (for L- and R-mode)  and Hall-Lorenz 
number ($\kappa_{1}/\sigma_{\text{Hall}}T$) 
(for L- and R-mode) for eB = 0.1$m_{\pi}^2$, 0.5 $m_{\pi}^2$
at $\mu$= 30 MeV
is plotted in Fig.\eqref{fig6} and \eqref{fig8}. 
Lorenz number in absence of magnetic field at 
finite chemical potential is shown in
 Fig.\eqref{lorenz_t}. Since, Lorenz and 
 Hall-Lorenz number is larger than unity 
 implying that the effect of thermal transport 
 coefficient is more pronounced than 
 charge transport coefficient, hence
suggesting that hot QCD matter is good 
conductor of heat than charge. It is evident 
that Lorenz and Hall Lorenz number is not 
constant with temperature. Lorenz number 
for L-mode increases with magnetic field
 whereas for R-mode decreases with magnetic 
 field. For L-mode, Lorenz number shows the increasing trend with temperature for $eB = 0.1 m_{\pi}^2$ whereas with further increase the magnetic field it start decreasing with temperature. This opposite behaviour with temperature is due to the difference in the increment of the ratio $(\kappa_{0}/\sigma_{{\text{Ohmic}}})$ at $eB = 0.1 m_{\pi}^2, 0.5 m_{\pi}^2$. Hall Lorenz 
 number for L-mode increases with magnetic 
 field whereas that for R-mode, it decreases with 
magnetic field similar to the previous case 
as shown in Fig.\eqref{fig6}. It increases with temperature for both modes.
Here, the behaviour of Lorenz number is 
found to be in contrast with the case 
of metals where it is roughly 
same in Drude model at temperature 
273 K and 373 K \cite{ashcroft}. Therefore, violation of Wiedemann-Franz law is observed.
\vspace{1cm}
\section{CONCLUSION}\label{conclusion}
In this work, we have studied the charge and 
heat transport coefficients in hot QCD 
matter in presence of weak magnetic field 
at finite chemical potential, where interactions
 have been incorporated through effective masses 
 using quasiparticle description. In weak magnetic field, we have found that the left- and right-handed 
 chiral modes of quarks get separated due to 
 difference in their mass and become non-degenerate 
 contrary to the strong magnetic field case. Another consequence of weak magnetic field also came into light in transport phenomena as generation of Hall effect. Transport coefficients adopts the tensorial structure where we get the non-vanishing transverse responses.
The diagonal elements of tensor structure of transport coefficients gives
longitudinal conductivity whereas off-diagonal 
elements represents their Hall counterparts. 
We have calculated the transport coefficients 
using the effective mass of quarks for 
left- (L) and right-handed (R) chiral modes separately and studied
 the effect of magnetic field and quark chemical potential
  on transport coefficients for both modes. We 
  studied the variation of $\sigma_{{\text{Ohmic}}}/T$ and $\sigma_{{\text{Hall}}}/T$ for L- and R- modes 
at different values of magnetic field and quark 
chemical potential with temperature. $\sigma_{{\text{Ohmic}}}/T$ for L-mode decreases with magnetic 
field whereas it increases with magnetic field for R-mode. The opposite behaviour with magnetic field for L- and R-mode in Ohmic conductivity is due to different values of effective quark mass for both modes. On the other hand, $\sigma_{{\text{Hall}}}/T$ for both modes increases
 with magnetic field. This is due the to direct 
 dependence of magnetic field on Hall conductivity. 
 Additionally, 
 both the conductivities for L- and R-mode 
 positively amplifies with quark chemical 
 potential. Hall conductivity vanishes at zero 
 quark chemical potential due to equal and 
opposite contribution of quarks and anti-quarks. 
Analogous to Ohmic and Hall conductivity, 
we have studied the thermal and Hall-type thermal 
conductivity for both modes. Since, gluons are not affected by magnetic field therefore thermal conductivity due to gluons is incorporated in longitudinal thermal conductivity. The Hall-type 
thermal conductivity is the manifestation of 
transverse temperature gradient under the 
action of Lorentz force. $\kappa_{0}/T$ for L- and 
R-mode shows mutually opposite behaviour with magnetic 
field which is again due to the different effective quark 
masses for left- and right-handed mode. $\kappa_{1}/T$ 
increases with 
magnetic field for both the modes similar to the Hall 
conductivity in charge transport. Both the 
conductivities record a drop in their values with increasing quark 
chemical potential. Moreover, $\kappa_{1}$ does not vanish at zero quark chemical potential due to the unequal contribution from quarks and anti-quarks in the same direction. In 
application of aforementioned conductivities, 
we have investigated the equilibrium property 
through Knudsen number ($\Omega_0$ and $\Omega_1$) 
where it is found to be less than unity ensuring 
the system to be in thermal equilibrium. The variation 
of Knudsen number with magnetic field is found to be 
closely related to the thermal and Hall-type thermal 
conductivity, as the specific heat at constant 
volume doesn't show significant change with 
 magnetic field. Further, 
the relative behaviour of charge and heat 
transport coefficients has been studied via 
Wiedemann-Franz law where Lorenz and Hall 
Lorenz number are found to be greater than unity, 
hence depicting that hot QCD matter is good 
conductor of heat. Moreover, Lorenz and Hall
 Lorenz number increases with magnetic 
 field for L-mode and decreases with magnetic 
 field for R-mode. Lorenz number for L-mode (at eB = 0.1 m$_{\pi}^2$) and for R-mode (at eB = 0.1 m$_{\pi}^2$, 0.5 m$_{\pi}^2$)
 increases with temperature. As we further increase 
 the magnetic field, Lorenz number for L-mode shows a decreasing trend 
with temperature. Since,
  Lorenz and Hall Lorenz number are not constant 
  with temperature, thereby violating the Wiedemann-Franz law.
\section*{Acknowledgements}
Pushpa Panday would like to acknowledge Debarshi 
Dey and Salman Ahamad Khan for  useful discussions.
\appendix
\appendixpage
\addappheadtotoc
\begin{appendices}
\renewcommand{\theequation}{A.\arabic{equation}}
\section{ CALCULATION OF STRUCTURE FUNCTIONS}\label{append}
Here, we will show the computation of structure functions from Eq.\eqref{A} to \eqref{D} in one-loop order for hot and weakly magnetized medium under HTL approximation.
 Since, trace of odd number of gamma matrices is zero, the Eq.\eqref{A} can be written as
\begin{equation}
	\mathcal{A} = \frac{1}{4}\frac{\text{Tr}(\Sigma_{0}\slashed{P})- (P.u)\text{Tr}(\Sigma_{0}\slashed{u})}{(P.u)^2-P^2},
\end{equation}
where, 
\begin{equation}
	\Sigma_{0} = g^2 C_F T\sum_{n}\int\frac{d^3 k}{(2\pi)^3}\gamma_{\mu}\frac{\slashed{K}}{K^2-m_f^2}\gamma^{\mu}\frac{1}{(P-K)^2}.
\end{equation}
Using the following two traces:
\begin{align}
	&\text{Tr}\left[\gamma_{\mu}\slashed{K}\gamma^{\mu}\slashed{P}\right] = -8K.P,\\
	&\text{Tr}\left[\gamma_{\mu}\slashed{K}\gamma^{\mu}\slashed{u}\right] = -8K.u,\\
\end{align}
 we obtain,
 \begin{equation}
 	\mathcal{A}(P) = \frac{1}{4\bf{|p|}^2}g^2 C_F\left[I_1(P)+I_2(P)\right],
 \end{equation}
 where $(P.u)^2-P^2 = \bf{|p|}^2$. We will use the frequency sum to evaluate $I_1(P)$ and $I_2(P)$ with $k_0 = i\omega_n$, $p_0 = i\omega$, $E_1 = \sqrt{k^2+m_{f0}^2}$ and $E_2 = \sqrt{(p-k)^2}$. The frequency sum for fermion-boson case is \cite{bellac}
 \begin{equation}
 	T\sum_{n}\tilde{\Delta}_{s1}(i\omega_n,E_1)\Delta_{s2}(i(\omega-\omega_n),E_2) =\sum_{s1,s2=\pm 1} -\frac{s_1 s_2}{4E_1E_2}\frac{\left(1-\tilde{f}(s_1E_1)+f(s_2E_2)\right)}{i\omega-s_1E_1-s_2E_2}.
 \end{equation}
The leading $T^2$ behaviour will come from $s_1 = -s_2 = 1$ with $E_1\approx k$ and $E_2= |\bf{p-k}|$. Defining light-like four-vector $\hat{K}=(-i,\bf{\hat{k}})$ and $\hat{K}^{\prime} = (-i,\bf{-\hat{k}})$, we have,
\begin{align}
&i\omega + E_1 - E_2\simeq i\omega + {\bf{p.\hat{k}}} = P.\hat{K},\\
&i\omega - E_1 + E_2\simeq i\omega - {\bf{p.\hat{k}}} = P.\hat{K^{\prime}},
\end{align}
 and using the angular integration under HTL approximation,
 \begin{equation}
 	\int\frac{d\Omega}{4\pi}\frac{\hat{K}.u}{P.\hat{K}} = \frac{1}{|\bf{p}|}Q_0\left(\frac{p_0}{|\bf{p}|}\right),
 \end{equation}
we get,
\begin{equation}
	\mathcal{A}(p_0,{|\bf{p}}|) = \frac{m_{th}^2}{{|\bf{p}}|^2}Q_1\left(\frac{p_0}{|\bf{p}|}\right).
\end{equation} 
Similarly, structure function $\mathcal{B}$ can be evaluated as
\begin{equation}
\mathcal{B}(p_0,{|\bf{p}}|) =- \frac{m_{th}^2}{{|\bf{p}|}^2}\left[\frac{p_0}{|\bf{p}|}Q_1\left(\frac{p_0}{|\bf{p}|}\right) - Q_0\left(\frac{p_0}{|\bf{p}|}\right)\right].
\end{equation} 
 Using Eq.\eqref{self_energy} in \eqref{C} and \eqref{D}, where the contribution from $\Sigma_{0}$ vanishes due to the trace of odd no. of gamma matrices and we get the non-vanishing contribution form $\Sigma_{1}$ only and hence we get,
 \begin{align}
 	&\mathcal{C}\left(p_0,{\bf{|p|}}\right)= -\frac{1}{4}\text{Tr}(\gamma_5\Sigma_{1}\slashed{u}),\\
 	& \mathcal{D}\left(p_0, {\bf{|p|}}\right)= \frac{1}{4}\text{Tr}(\gamma_5\Sigma_{1}\slashed{b}).
 \end{align}
 Using the following two traces
\begin{align}
	&\text{Tr}\left[\gamma_5\gamma_{\mu}\gamma_5\left[(K.b)\slashed{u}-(K.u)\slashed{b}\right]\gamma^{\mu}\slashed{u}\right] = 8(K.b),\\
	&\text{Tr}\left[\gamma_5\gamma_{\mu}\gamma_5\left[(K.b)\slashed{u}-(K.u)\slashed{b}\right]\gamma^{\mu}\slashed{b}\right] = 8(K.u),
\end{align} 
 we obtain,
 \begin{align}
 &	\mathcal{C} = \frac{g^2 C_F|q_f B|}{4}T\sum_{n}\int\frac{d^3k}{(2\pi)^3}\frac{8 (K.b)}{(K^2-m_{f0}^2)^2(P-K)^2},\\
 &	\mathcal{D} = -\frac{g^2 C_F|q_f B|}{4}T\sum_{n}\int\frac{d^3k}{(2\pi)^3}\frac{8 (K.u)}{(K^2-m_{f0}^2)^2(P-K)^2},
 \end{align}
 which in turn requires the calculation of frequency sum \cite{ayala}
 \begin{align}
 &	Y = T\sum_{n}\Delta_F^2(K)\Delta_B(P-K), \\ \nonumber
 &	= \left(\frac{-\partial}{\partial m_{f0}^2}\right) T\sum_{n}\Delta_{F}(K)\Delta_{B}(P-K),
 \end{align}
 where,
 \begin{equation}
 	T\sum_{n}\Delta_{F}(K)\Delta_{B}(P-K) = \sum_{s1,s2=\pm 1} -\frac{s_1 s_2}{4E_1E_2}\frac{\left(1-\tilde{f}(s_1E_1)+f(s_2E_2)\right)}{i\omega-s_1E_1-s_2E_2} .
 \end{equation}
 For $s_1 = -s_2 = 1$, we get,
 \begin{align}
 &	\mathcal{C} = \frac{4g^2C_F|q_fB|}{16\pi^2}\left(\frac{\pi T}{2m_{f0}} -\text{ln} 2 + \frac{7\mu^2\zeta(3)}{8\pi^2T^2}\right)\left[\frac{-p_z}{{\bf|p|}^2}Q_1\left(\frac{p_0}{|\bf{p}|}\right)\right],\\
 &	\mathcal{D} = -\frac{4g^2C_F|q_fB|}{16\pi^2}\left(\frac{\pi T}{2m_{f0}} -\text{ln} 2 + \frac{7\mu^2\zeta(3)}{8\pi^2T^2}\right)\left[\frac{1}{{\bf|p|}}Q_0\left(\frac{p_0}{|\bf{p}|}\right)\right].\\
\end{align}

\end{appendices}


\end{document}